\documentclass[twocolumn]{aastex62}

\usepackage[caption=false]{subfig}
\usepackage{float}
\usepackage{graphicx}	
\usepackage{amsmath}	
\usepackage{amsfonts}	
\usepackage{amssymb}
\usepackage{xcolor}

\newcommand{\msun}{\mbox{$M_\odot$}} 

\newcommand{\rsun}{\mbox{$R_\odot$}}

\newcommand{\mjup}{M$_{\rm J}$}
\newcommand{\rjup}{$R_{\rm J}$}

\newcommand{\ms}{\mbox{m s$^{-1}$}}

\newcommand{\kms}{\mbox{km s$^{-1}$}}

\graphicspath{{./}}

\received{}
\accepted{}

\shorttitle{Iron in TOI-1518b}
\shortauthors{Cabot et al.}

\begin{document}

\title{TOI-1518b: A Misaligned Ultra-hot Jupiter with Iron in its Atmosphere}

\correspondingauthor{Samuel H. C. Cabot}
\email{sam.cabot@yale.edu}

\author[0000-0001-9749-6150]{Samuel H. C. Cabot}
\affil{Yale University, 52 Hillhouse Avenue, New Haven, CT 06511, USA}

\author[0000-0003-3355-1223]{Aaron Bello-Arufe}
\affil{National Space Institute, Technical University of Denmark, Elektrovej, DK-2800 Kgs. Lyngby, Denmark}

\author[0000-0002-6907-4476]{João M. Mendonça}
\affil{National Space Institute, Technical University of Denmark, Elektrovej, DK-2800 Kgs. Lyngby, Denmark}

\author[0000-0003-1001-0707]{Ren\'e Tronsgaard}
\affil{National Space Institute, Technical University of Denmark, Elektrovej, DK-2800 Kgs. Lyngby, Denmark}

\author[0000-0001-9665-8429]{Ian Wong}\altaffiliation{51 Pegasi b Fellow}
\affil{Department of Earth, Atmospheric and Planetary Sciences, Massachusetts Institute of Technology, Cambridge, MA 02139, USA}

\author[0000-0002-4891-3517]{George Zhou}
\affil{Center for Astrophysics $\vert$ Harvard \& Smithsonian, 60 Garden Street, Cambridge, MA 02138, USA}

\author[0000-0003-1605-5666]{Lars A. Buchhave}
\affil{National Space Institute, Technical University of Denmark, Elektrovej, DK-2800 Kgs. Lyngby, Denmark}

\author[0000-0003-2221-0861]{Debra A. Fischer}
\affil{Yale University, 52 Hillhouse Avenue, New Haven, CT 06511, USA}

\author[0000-0002-3481-9052]{Keivan G. Stassun}
\affil{Vanderbilt University, Department of Physics \& Astronomy, 6301 Stevenson Center Ln., Nashville, TN 37235, USA}

\author[0000-0002-0865-3650]{Victoria Antoci}
\affil{National Space Institute, Technical University of Denmark, Elektrovej, DK-2800 Kgs. Lyngby, Denmark}
\affil{Stellar Astrophysics Centre, Department of Physics and Astronomy, Aarhus University, Ny Munkegade 120, DK-8000 Aarhus C, Denmark}

\author[0000-0002-2970-0532]{David Baker} 
\affil{Physics Department, Austin College, Sherman, TX 75090, USA}

\author[0000-0003-3469-0989]{Alexander A.\ Belinski}
\affiliation{Sternberg Astronomical Institute, M.V. Lomonosov Moscow State University, 13, Universitetskij pr., 119234, Moscow, Russia}

\author[0000-0001-5578-1498]{Bj{\"o}rn~Benneke}
\affiliation{Department of Physics and Institute for Research on Exoplanets, Universit{\'e} de Montr{\'e}al, Montreal, QC, Canada}

\author[0000-0002-0514-5538]{Luke G. Bouma}
\affiliation{Department of Astrophysical Sciences, Princeton University, NJ 08544, USA}

\author[0000-0002-8035-4778]{Jessie L. Christiansen}
\affiliation{NASA Exoplanet Science Institute – Caltech/IPAC Pasadena, CA 91125 USA}

\author[0000-0001-6588-9574]{Karen A.\ Collins}
\affil{Center for Astrophysics $\vert$ Harvard \& Smithsonian, 60 Garden Street, Cambridge, MA 02138, USA}

\author{Maria V. Goliguzova}
\affiliation{Sternberg Astronomical Institute, M.V. Lomonosov Moscow State University, 13, Universitetskij pr., 119234, Moscow, Russia}

\author[0000-0001-8072-0590]{Simone Hagey} 
\affil{University of Saskatchewan, Saskatchewan, Canada}

\author[0000-0002-4715-9460]{Jon M. Jenkins}
\affil{NASA Ames Research Center, Moffett Field, CA, 94035}

\author[0000-0002-4625-7333]{Eric L. N. Jensen} 
\affiliation{Dept.\ of Physics \& Astronomy, Swarthmore College, Swarthmore PA 19081, USA} 

\author{Richard C. Kidwell Jr}
\affil{Space Telescope Science Institute, Baltimore, MD, USA}

\author{Didier Laloum}
\affil{Société Astronomique de France, 3 Rue Beethoven, 75016 Paris, France}

\author[0000-0001-8879-7138]{Bob Massey}
\affil{Villa '39 Observatory, Landers, CA 92285, USA}

\author[0000-0001-9504-1486]{Kim K. McLeod}
\affil{Department of Astronomy, Wellesley College, Wellesley, MA 02481, USA}

\author[0000-0001-9911-7388]{David W. Latham}
\affiliation{Center for Astrophysics $\vert$ Harvard \& Smithsonian, 60 Garden Street, Cambridge, MA 02138, USA}

\author{Edward~H.~Morgan} 
\affiliation{Department of Physics and Kavli Institute for Astrophysics and Space Research, Massachusetts Institute of Technology, Cambridge, MA 02139, USA}

\author[0000-0003-2058-6662]{George Ricker}
\affiliation{Department of Physics and Kavli Institute for Astrophysics and Space Research, Massachusetts Institute of Technology, Cambridge, MA 02139, USA}

\author[0000-0003-1713-3208]{Boris S.\ Safonov}
\affiliation{Sternberg Astronomical Institute, M.V. Lomonosov Moscow State University, 13, Universitetskij pr., 119234, Moscow, Russia}

\author[0000-0001-5347-7062]{Joshua~E.~Schlieder}
\affiliation{NASA Goddard Space Flight Center, 8800 Greenbelt Rd, Greenbelt, MD 20771, USA}

\author[0000-0002-6892-6948]{Sara Seager}
\affiliation{Department of Physics and Kavli Institute for Astrophysics and Space Research, Massachusetts Institute of Technology, Cambridge, MA 02139, USA}
\affiliation{Department of Earth, Atmospheric and Planetary Sciences, Massachusetts Institute of Technology, Cambridge, MA 02139, USA}
\affiliation{Department of Aeronautics and Astronautics, MIT, 77 Massachusetts Avenue, Cambridge, MA 02139, USA}

\author[0000-0002-1836-3120]{Avi Shporer}
\affiliation{Department of Physics and Kavli Institute for Astrophysics and Space Research, Massachusetts Institute of Technology, Cambridge, MA 02139, USA}

\author[0000-0002-6148-7903]{Jeffrey C. Smith}
\affil{NASA Ames Research Center, Moffett Field, CA, 94035}
\affil{SETI Institute, Mountain View, CA  94043, USA}

\author{Gregor Srdoc}
\affil{Kotizarovci Observatory, Sarsoni 90, 51216 Viskovo, Croatia}

\author{Ivan A. Strakhov}
\affiliation{Sternberg Astronomical Institute, M.V. Lomonosov Moscow State University, 13, Universitetskij pr., 119234, Moscow, Russia}

\author[0000-0002-5286-0251]{Guillermo Torres}
\affiliation{Center for Astrophysics $\vert$ Harvard \& Smithsonian, 60 Garden Street, Cambridge, MA 02138, USA}

\author[0000-0002-6778-7552]{Joseph D. Twicken}
\affil{NASA Ames Research Center, Moffett Field, CA, 94035}
\affil{SETI Institute, Mountain View, CA  94043, USA}

\author[0000-0001-6763-6562]{Roland Vanderspek}
\affiliation{Department of Physics and Kavli Institute for Astrophysics and Space Research, Massachusetts Institute of Technology, Cambridge, MA 02139, USA}

\author{Michael~Vezie}
\affiliation{Department of Physics and Kavli Institute for Astrophysics and Space Research, Massachusetts Institute of Technology, Cambridge, MA 02139, USA}

\author[0000-0002-4265-047X]{Joshua N.\ Winn}
\affiliation{Department of Astrophysical Sciences, Princeton University, NJ 08544, USA}



\begin{abstract}

We present the discovery of TOI-1518b --- an ultra-hot Jupiter orbiting a bright star ($V = 8.95$). The transiting planet is confirmed using high-resolution optical transmission spectra from EXPRES. It is inflated, with {$R_p = 1.875\pm0.053\,R_{\rm J}$}, and exhibits several interesting properties, including a misaligned orbit (${240.34^{+0.93}_{-0.98}}$ degrees) and nearly grazing transit {($b =0.9036^{+0.0061}_{-0.0053}$)}. The planet orbits a fast-rotating F0 host star ($T_{\mathrm{eff}} \simeq 7300$~K) in 1.9 days and experiences intense irradiation. Notably, the TESS data show a clear secondary eclipse with a depth of {$364\pm28$~ppm} and a significant phase curve signal, from which we obtain a relative day--night planetary flux difference of roughly 320~ppm and a {5.2$\sigma$} detection of ellipsoidal distortion on the host star. Prompted by recent detections of atomic and ionized species in ultra-hot Jupiter atmospheres, we conduct an atmospheric cross-correlation analysis. We detect neutral iron (${5.2\sigma}$), at ${K_p = {157^{+68}_{-44}}}$ \kms\ and ${V_{\rm sys} = {-16^{+2}_{-4}}}$ \kms, {adding} another object to the small sample of highly irradiated gas-giant planets with Fe detections in transmission. Detections so far favor particularly inflated gas giants with radii $\gtrsim 1.78$\rjup; {although this may be due to observational bias.} With an equilibrium temperature of {$T_{\rm eq}=2492\pm38$~K} and a measured dayside brightness temperature of {$3237\pm59$~K} (assuming zero geometric albedo), TOI-1518b is a promising candidate for future emission spectroscopy to probe for a thermal inversion.

\end{abstract}


\keywords{Exoplanets --- Hot Jupiters --- Exoplanet atmospheres --- Spectroscopy}

\section{Introduction} \label{sec:intro}

Transiting exoplanets --- those that pass directly between their host stars and an observer --- offer a wealth of information about their systems. The transit itself is detectable through the minuscule fraction of starlight occulted by the planet, which is well within the sensitivity of many current ground- and space-based telescopes. The {\it Kepler} \citep{Borucki2010} and K2 \citep{Howell2014} missions together yielded thousands of transiting exoplanet candidates, some of which are among the most notable and well-characterized to date. Planets found by surveys such as HATNet \citep{Bakos2004}, KELT \citep{Pepper2007}, and WASP \citep{Pollacco2006} orbit some of the brightest stars, and have hence been popular targets for atmospheric characterization. Today, the frontier lies with the {\it Transiting Exoplanet Survey Satellite} (TESS; \citealt{Ricker2014}), which is searching for planets transiting bright stars across the entire sky.

The science drivers behind exoplanet transit observations are several-fold. Newly discovered systems improve our baseline understanding of exoplanet populations and distributions \citep{Howard2012, Fressin2013, Fulton2017}, and how their properties may be linked to system architecture and formation scenarios \citep{Lissauer2011, Fabrycky2014, Millholland2017, Weiss2018}. The presence of additional, non-transiting exoplanets can be inferred from transit timing perturbations \citep{Holman2005, Ballard2011}. The host star's obliquity can be probed by the Rossiter-McLaughlin effect \citep{Winn2010, Triaud2018}, which also relates to formation pathways \citep[][and references therein]{Dawson2018}. Finally, transits enable the study of exoplanet atmospheres based on the excess absorption of starlight from high-altitude species \citep[e.g.][]{Seager1998, Charbonneau2002, Snellen2010, Sing2016}. These latter investigations require dedicated spectroscopic followup.

In this study, we present the confirmation of the TESS transiting planet candidate TOI-1518b, a highly irradiated gas giant planet possessing iron vapor in its atmosphere. Of exoplanets discovered by TESS, this is the first high-resolution detection of an atmospheric species. Several TESS candidates have been confirmed as hot Jupiters so far, including HD 202772Ab \citep{Wang2019}, HD 2685b \citep{Jones2019}, TOI-150b \citep{Canas2019}, HD 271181b \citep{Kossakowski2019}, TOI-172b \citep{Rodriguez2019}, TOI-564b, and TOI-905b \citep{Davis2019}. However, TOI-1518b is unique due to its close-in orbit (1.9 day period) and high level of irradiation from its F-type host star. 

The new planet falls within the category of ultra-hot Jupiters (UHJs), which have equilibrium temperatures exceeding 2000 K \citep{Fortney2008, Parmentier2018}. Many UHJs contain vaporized metals, both neutral and ionized, in their upper atmospheres \citep[e.g.][]{Hoeijmakers2018, Casasayas-Barris2018}. These metals and molecules containing them are recognized as strong sources of opacity in the optical and near-ultraviolet regions \citep{Fortney2008,Lothringer2020}. UHJs often exhibit thermal inversions \citep{Haynes2015,Evans2017}; however, the exact species responsible for the inversions are debated \citep{Fortney2008, Lothringer2018, Gandhi2019}. High-resolution spectroscopy has become a common method for detecting important species in UHJ atmospheres and also serves as a means of probing winds \citep{Louden2015, CasasayasBarris2019} and extended atmospheres \citep{Yan2018}.  

Our paper is organized as follows. In Section~\ref{sec:obs} we analyze the TESS photometry of TOI-1518. We reproduce the detection of a planet candidate, obtain constraints on its orbital parameters, and report a robust detection of the secondary eclipse and phase-curve modulations. We also present high-resolution spectroscopic observations of the system during transit. This spectroscopic transit is analyzed in Section~\ref{sec:cca}, from which we measure the Rossiter-Mclaughlin effect and obtain further constraints on the orbit and host star. Section~\ref{sec:cca} also includes a review of the cross-correlation method for atmospheric characterization, the results of which are presented in Section~\ref{sec:atm}. Finally, we discuss TOI-1518b in the context of previously studied UHJs in Section~\ref{sec:discussion}.
 
\section{Observations and System Characterization} \label{sec:obs}

\begin{figure*}
    \centering
    \includegraphics[width=\linewidth]{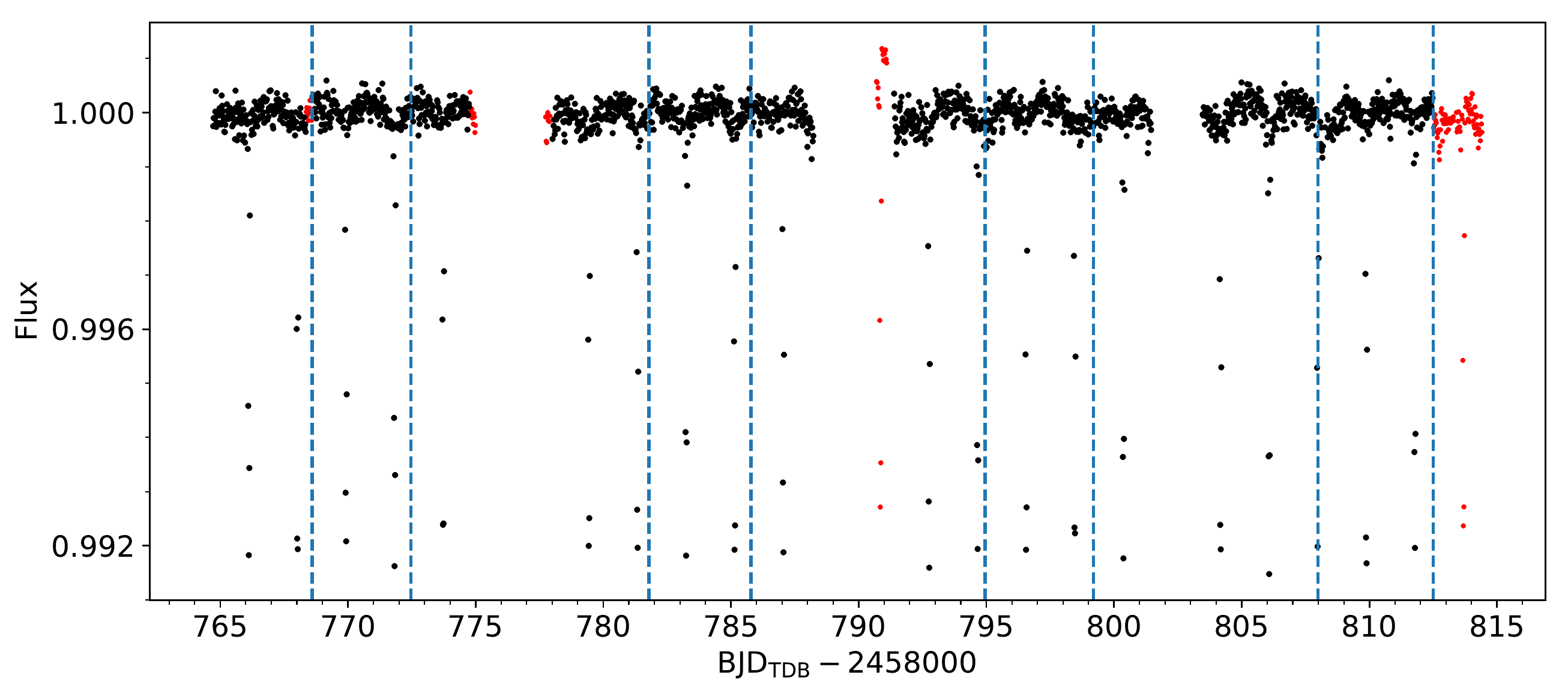}
    \caption{The normalized Presearch Data Conditioning Simple Aperture Photometry (PDCSAP) light curve of TOI-1518 generated by the SPOC pipeline. The scheduled momentum dumps are indicated by the vertical blue dashed lines. The red points denote flux ramps and regions of severe systematics that were trimmed prior to our light curve fits. The orbital phase curve modulations are discernible in the raw photometry.}
    \label{fig:tessraw}
\end{figure*}

This section describes our analysis of the available TESS photometry of TOI-1518, as well as high-resolution optical spectra of the system. We measure the system parameters by simultaneously modeling the transit, secondary eclipse, and full-orbit phase curve. We also fit spectral lines to determine properties of the star. As detailed below, radial velocity (RV) measurements of the system provide some broad constraints. However, the deduced parameters have large uncertainties owing to the rapid rotation speed of the star. 

\subsection{TESS Photometry}
\label{subsec:tess}

The star TIC 427761355 (also designated as BD+66 1610) was observed by Camera 3 of the TESS instrument during Sectors 17 and 18 (UT 2019 Oct 7 to Nov 27). The Quick Look Pipeline \citep[QLP; ][]{Huang2020} detected a likely transit signal in the photometry and flagged the companion as a candidate transiting exoplanet with parameters characteristic of a close-in hot Jupiter. The system was released as a TESS Object of Interest (TOI) with the designation TOI-1518. The full-frame images (FFIs) were processed by the Science Processing Operations Center \citep[SPOC; ][]{jenkinsSPOC2016} and made publicly available on the Mikulski Archive for Space Telescopes (MAST)\footnote{https://mast.stsci.edu/}. 

We obtained the TESS-SPOC HLSP light curves \citep{Caldwell2020} for TOI-1518 from MAST. The SPOC data includes two versions of the photometry at the standard 30-minute cadence: (1) the Simple Aperture Photometry (SAP) light curve, i.e., the raw photometry extracted from the SPOC pipeline-derived photometric aperture \citep{twicken:PA2010SPIE, morris2020}, and (2) the Presearch Data Conditioning SAP (PDCSAP) light curve, which has been corrected for common-mode systematics trends shared by other sources on the detector {(i.e., co-trending basis vectors, or CBVs)}, while preserving the key astrophysical signals of interest \citep{Stumpe2012, Stumpe2014, Smith2012}. 

The PDCSAP light curve is considerably cleaner than the SAP photometry, and in this paper, we present the analysis of the PDCSAP light curve. {For completeness, we carried out an analogous analysis of the SAP light curve. Systematics were modeled using linear combinations of the CBVs, similar to the detrending methodology in the SPOC pipeline. We obtained results that are statistically consistent with the main PDCSAP-derived values to within $1\sigma$. However, there were residual long-term systematics trends even after detrending with the CBVs, which led to a roughly $10\%$ increase in residual scatter from the best-fit light-curve model when compared to the PDCSAP analysis.}

Our analysis methodology closely mirrors the techniques utilized in the extensive previous work on TESS phase curves \citep[e.g.,][]{shporer2019,wong2020wasp19,wong2020year1,wong2020kelt9}; consult those references for a detailed description of the data processing and light-curve fitting. The full PDCSAP light curve of TOI-1518 is shown in Figure~\ref{fig:tessraw}. Each TESS Sector consists of two spacecraft orbits, separated by a pause in science observations for data downlink. Momentum dumps are scheduled during each spacecraft orbit to reset the onboard reaction wheels. In Sectors 17 and 18, these occurred twice per spacecraft orbit and are indicated in the plot by vertical blue dashed lines. The momentum dumps induce small discontinuities in the photometry, as well as occasional short-term flux ramps. We therefore divide the light curve into individual segments separated by the momentum dumps and model the remaining systematics within each segment separately. Significant ramps are trimmed prior to the final fit; the trimmed points are shown in Figure~\ref{fig:tessraw} in red. The last data segment of Sector 18 is not included in our analysis due to severe residual systematics. We also apply a 16-point-wide moving median filter to the light curve after masking the transits and remove $3\sigma$ outliers. The final light curve contains 1,845 points, divided among 11 segments.

Visual inspection of Figure~\ref{fig:tessraw} reveals coherent flux modulations synchronized to the planet's orbit, indicative of a phase curve. {To examine the harmonic content of the TESS photometry in more detail, we trim the transits and secondary eclipses from the light curve (after correcting for instrumental systematics; see Section~\ref{subsec:phasecurve}) and generate the Lomb--Scargle periodogram. The result is plotted in Figure~\ref{fig:ls}. We find a very strong signal at the orbital frequency, as well as another significant periodicity at the first harmonic of the orbital period (i.e., two maxima per orbital period).} 

The phase curve of a star--planet system formally contains contributions from both the planet and the host star (see review in \citealt{shporer2017}). Close-in exoplanets are tidally-locked, with fixed dayside and nightside hemispheres; as the planet rotates, the viewing geometry changes, resulting in a periodic modulation of the observed atmospheric flux that varies as the cosine of the orbital phase. Massive orbiting companions can also raise a tidal bulge on the host star's surface, resulting in a periodic flux modulation that comes to maximum at quadrature (i.e., a signal {with a leading-order term} at the first harmonic of the cosine); this is typically referred to as ellipsoidal distortion. Lastly, the mutual star--planet gravitational interaction causes Doppler shifting of the star's spectrum, producing a modulation in the total system flux within the bandpass that can sometimes by detected in visible-light photometry. This so-called Doppler boosting signal has the same phase alignment as the RV signal, i.e., the sine of the orbital phase. 

\subsection{{Full-orbit Phase-curve Model}}
\label{subsec:phasecurve}

We fit the full-orbit phase curve with a composite flux model for the planet $\psi_{p}$ and the star $\psi_\star$ \citep[e.g.,][]{wong2020year1, wong2020kelt9,wong2021year2}:
\begin{align}
\label{planet}\psi_{p}(t) &= f_{p} - A_{\mathrm{atm}} \cos(\phi+\delta),\\
\label{star}\psi_\star(t) &= 1-A_{\mathrm{ellip}}\cos(2\phi)+A_{\mathrm{Dopp}}\sin(\phi).
\end{align}
Here, $A_{\mathrm{atm}}$, $A_{\mathrm{ellip}}$, and $A_{\mathrm{Dopp}}$ indicate the semiamplitudes of the planet's atmospheric brightness modulation, the star's ellipsoidal distortion signal, and the Doppler boosting, respectively; the signs are assigned so that the measured amplitudes are positive under normal circumstances. The variables $f_{p}$ and $\delta$ signify the average relative brightness of the planet across its orbit and the phase shift in the planet's phase curve, respectively.

\begin{figure}[t]
    \centering
    \includegraphics[width=\linewidth]{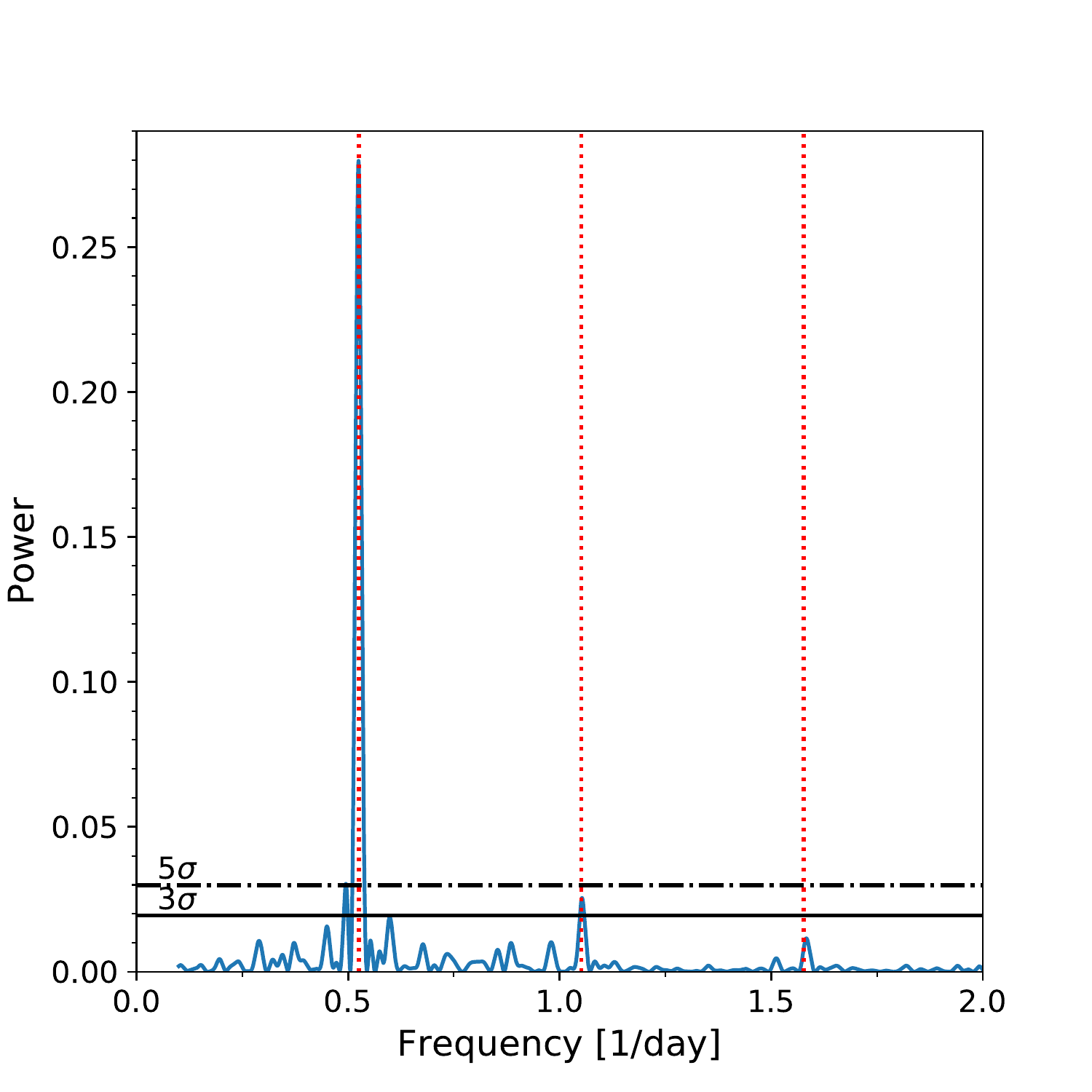}
    \caption{{Lomb--Scargle periodogram of the detrended TESS PDCSAP light curve of TOI-1518, with the transits and secondary eclipses removed. Significance thresholds are indicated by the horizontal lines. The red vertical lines denote the first three harmonics of the orbital period. There are clear signals at the orbital frequency and at the first harmonic, corresponding to the planetary atmospheric brightness modulation and stellar ellipsoidal distortion, respectively.}}
    \label{fig:ls}
\end{figure}

We note that the stellar ellipsoidal distortion signal contains additional higher-order terms \citep[e.g.,][]{morris1985,shporer2017}. The second-highest amplitude is expected at the second harmonic of the cosine (i.e., $\cos(6\phi)$). However, there is no significant power {precisely} at that harmonic in the Lomb--Scargle periodogram (Figure~\ref{fig:ls}){; the weak signal around 1.6~d$^{-1}$ is centered at a slightly higher frequency than the second harmonic and is likely attributable to low-level residual systematics in the light curve. Indeed, when} fitting for the second-harmonic amplitudes in the light-curve analysis, we do not measure any significantly nonzero amplitudes. Therefore, we do not include any higher-order terms of the ellipsoidal distortion when generating the final set of phase-curve fit results.

The transits and secondary eclipse light curves ($\lambda_t$ and $\lambda_e$) are modeled using \textsf{batman} \citep{batman}. The secondary eclipse depth (i.e., total dayside hemisphere flux) is related to the phase-curve parameters via the expression $D_{d}=f_{p} - A_{\mathrm{atm}} \cos(\pi+\delta)$. Likewise, the hemisphere-averaged nightside flux is given by $D_{n}=f_{p} - A_{\mathrm{atm}} \cos(\delta)$. {To accurately model the 30-minute exposures during transit and secondary eclipse, we use an oversampling factor of 60, i.e., averaging the flux from 30-second subexposures at each timestamp.}

Any remaining systematics trends in each light curve segment $k$ are detrended using generalized polynomials in time:
\begin{equation}\label{systematics}
    S_N^{\lbrace k\rbrace}(t) = \sum\limits_{j=0}^{N}c_j^{\lbrace k\rbrace}(t-t_0)^j,
\end{equation}
where $t_0$ is the first timestamp of the segment, and $N$ is the order of the detrending polynomial, which in the final joint fit is set to the order that minimizes the Bayesian information criterion (BIC) for each segment. The optimal polynomial orders for the 11 light-curve segments included in our analysis are 2, 0, 0, 1, 0, 3, 3, 2, 3, 1, and 3. The total astrophysical-plus-systematics light-curve model, normalized to unity, is 

\begin{equation}
    \label{total}F(t) = \frac{\psi_\star(t)\lambda_t(t)+\psi_{p}(t)\lambda_e(t)}{1+f_p} \times S_N^{\lbrace k\rbrace}(t).
\end{equation}
    
{To obtain an initial set of results from the TESS photometry, we} jointly fit all 11 light-curve segments using the affine-invariant Markov chain Monte Carlo (MCMC) sampler \textsf{emcee} \citep{emcee}. The free astrophysical parameters in our fit that are unconstrained by any priors include the transit ephemeris (mid-transit time $T_{c}$ and orbital period $P$), transit shape parameters (impact parameter $b$ and scaled semimajor axis $a/R_\star$), planet--star radius ratio $R_{p}/R_\star$, and the phase-curve parameters. The predicted Doppler boosting amplitude assuming the RV-derived mass (see Section~\ref{subsec:fies}) is roughly 2~ppm --- significantly smaller than the uncertainties on the phase-curve amplitudes. Therefore, we do not fit the Doppler signal, while allowing $f_{p}$, $A_{\mathrm{atm}}$, $A_{\mathrm{ellip}}$, and $\delta$ to vary. We also include a uniform per-point uncertainty parameter $\sigma_{k}$ for each light-curve segment as a free parameter in order to ensure a reduced $\chi^{2}$ value of one and retrieve realistic uncertainties on the astrophysical parameters. The median values of $\sigma_{k}$ range from 147 to 190~ppm across the 11 segments. 

The low cadence of the photometry and the grazing nature of the planetary transit mean that the stellar limb darkening is not well constrained by the light curve. We employ the standard quadratic limb-darkening law and apply Gaussian priors to each coefficient. The median values are set to the values from \citet{claret2018}, interpolated for the measured stellar parameters (see Section~\ref{subsec:specmod}) of TOI-1518: $u_1 = 0.28$ and $u_2=0.23$; the width of the Gaussian is generously set to 0.05, which is several times larger than the corresponding range of coefficient values spanned by the stellar parameter uncertainty regions.

{From our preliminary fit to the full TESS light curve, we find that t}he transit is grazing, corresponding to a planet--star radius ratio of $R_{p}/R_\star=0.0987\pm0.0017$ and well-constrained {transit-shape parameters: $b=0.9103\pm0.0065$ and $a/R_\star=4.231\pm0.064$}. We detect the secondary eclipse with a depth of {$\sim$380~ppm} and a significant atmospheric phase-curve modulation with a semiamplitude of {roughly 160}~ppm. There is a nearly $5\sigma$ detection of the ellipsoidal distortion signal from the host star, with a semiamplitude of {around 30}~ppm.

To probe for deviations from a circular orbit, we also carry out a separate light-curve fit with the orbital eccentricity $e$ and argument of periastron $\omega$ as additional free parameters. From the photometry, the orbital eccentricity is mostly constrained by the timing of the secondary eclipse relative to the mid-transit time and, to a much lesser extent, the relative durations of the transit and secondary eclipse. We obtain a tight $2\sigma$ upper limit of $e<0.01$ (formally, $e=0.0031_{-0.0022}^{+0.0047}$); the inclusion of $e$ and $\omega$ as free parameters is strongly disfavored by the Bayesian Information Criterion ($\Delta\mathrm{BIC}=16$). The corresponding $e\cos\omega$ and $e\sin\omega$ values, which relate to offsets in the secondary eclipse timing and duration, respectively, are $0.0007_{-0.0012}^{+0.0016}$ and $-0.0005_{-0.0061}^{+0.0030}$. We therefore conclude that the orbit of TOI-1518b is consistent with circular.

Due to the relatively short timespan contained within each segment, there is a possibility of small correlations between the coefficients in the detrending polynomials and the phase-curve parameters. To examine the effect of our choice of polynomial orders, we experiment with allowing only polynomials up to first order (i.e., no curvature in the systematics model). The results from the corresponding joint fit agree well with the {aforementioned values}. In particular, the measured secondary eclipse depth, atmospheric brightness modulation amplitude, and stellar ellipsoidal distortion amplitude are statistically consistent at much better than the $1\sigma$ level. Therefore, we conclude that the optimized polynomial orders listed above, which include orders as high as 3, do not bias the astrophysical parameters in any significant way.

\subsection{Ground-based Light Curves}\label{subsec:groundlc}

We acquired ground-based time-series follow-up photometry of TOI-1518 as part of the TESS Follow-up Observing Program (TFOP)\footnote{https://tess.mit.edu/followup}. We used the \textsf{TESS Transit Finder}, which is a customized version of the \textsf{Tapir} software package \citep{Jensen:2013}, to schedule our transit observations. The photometric data were extracted using \textsf{AstroImageJ} \citep{Collins:2017}.

A full transit was observed from Adams Observatory at the Austin College (Sherman, TX, USA) 0.6\,m telescope on UT 2020 January 5 in I-band {($\lambda_{\rm eff} = 806$\,nm)}. A nearly full in-transit portion of a transit was observed from the Whitin Observatory (Wellesley, MA, USA) 0.7\,m telescope on UT 2020 January 6 in Sloan $g'$-band {($\lambda_{\rm eff} = 475$\,nm)}. A full transit was observed from the private Observatory of the Mount (Saint-Pierre-du-Mont, France) 0.2\,m telescope on UT 2020 January 08 in R-band {($\lambda_{\rm eff} = 647$\,nm)}. A full transit was observed from the Kotizarovci Observatory (Viskovo, Croatia) 0.3\,m telescope on UT 2020 January 12 in the Baader R 610\,nm longpass band ($R_{\rm long}$; $\lambda_{\rm cut-on} = 610$\,nm). A full transit was observed from the Villa\,'39 observatory (Landers, CA, USA) 0.36\,m telescope on UT 2020 January 24 in B-band {($\lambda_{\rm eff} = 442$\,nm)}. An egress was observed from the University of Saskatchewan Observatory (Saskatoon, SK, Canada) 0.3\,m telescope on UT 2020 March 23 {using an Astrodon Clear with Blue Blocking Filter (CBB; $\lambda_{\rm cut-on} = 500$\,nm)}. {The light-curve data are available at ExoFOP-TESS.\footnote{https://exofop.ipac.caltech.edu/tess} The raw ground-based transit light curves are shown in the Appendix.}

The follow-up light curves confirm that the TESS-detected event occurs on target relative to known Gaia stars. {We analyze the five transit observations with full event coverage (i.e., excluding the UT 2020 March 23 egress-only light curve) by fitting each time series with \textsf{batman}. The mid-transit time, orbital period, impact parameter, and scaled semimajor axis are constrained by Gaussian priors based on the results of the TESS phase-curve fit (Section~\ref{subsec:phasecurve}). Similar to our treatment of the TESS-band transit modeling, the limb-darkening coefficients are constrained by priors derived by interpolating the tabulated values in \citet{claret2013} for the appropriate bandpass to the measured stellar parameters and uniformly applying a Gaussian width of 0.05. In the case of the non-standard $R_{\rm long}$ filter used for the UT 2020 January 12 observation, we approximate the bandpass with the Cousins I-band. The systematics trends in every transit light curve are modeled as a linear combination of the airmass and the width of the target's point-spread-function, along with a constant offset for normalization.}

{The comparatively low signal-to-noise of the ground-based transit datasets translates to large relative uncertainties on the measured transit depth, exceeding 10\% across all five visits. Nevertheless, we obtain $R_p/R_\star$ values that are consistent with the measurement from fitting the TESS light curve alone at better than the $2\sigma$ level. Similarly, the five ground-based transit depths are mutually consistent to within $2\sigma$, indicating an achromatic transit.}

\subsection{{Joint Photometric Analysis}}\label{subsec:joint}
{To leverage the additional time baseline and complementary constraints on transit geometry provided by the follow-up transit light curves, we carry out a joint analysis of the TESS photometry and ground-based observations. The orbital ephemeris, transit-shape, and phase-curve parameters are allowed to freely vary, while the limb-darkening coefficients remain constrained by the previously-defined priors. The astrophysical light curve and instrumental systematics are simultaneously modeled for all six datasets in the MCMC analysis.}

\setlength{\tabcolsep}{5pt}
\begin{table*}
\centering
\begin{tabular}{l  c c r }
\hline
QLP/Atlas Parameters & Symbol & Units & Value \\
\hline
  Right Ascension  & {RA}        & ---  & 23 29 04.224                \\
  Declination      & {Dec}        & ---  & +67 02 05.377                \\
  V-band Magnitude      & $V$      & mag.  & $8.952$           \\
\hline 
\hline 
Transit and Orbital Parameters  \\
\hline 
  Orbital Period            & $P$              & days  & $1.902603 \pm 0.000011$   \\
  Mid-transit Time & $T_c$          & BJD$_{\mathrm{TDB}}$   & $2458787.049255 \pm 0.000094$ \\
  Radius Ratio      & $R_p/R_\star$        & ---     & $0.0988^{+0.0015}_{-0.0012}$ \\
  Impact Parameter  & $b$              & ---     & $0.9036_{-0.0053}^{+0.0061}$ \\
  Scaled Semimajor Axis   & $a/R_\star$   & ---     & $4.291_{-0.061}^{+0.057}$ \\
  Orbital Eccentricity & $e$ & --- & $<0.01$ ($2\sigma$) \\
  Orbital Inclination$^{\dag}$ & $i_p$            & deg.  & $77.84^{+0.23}_{-0.26}$ \\
  \hline 
\hline 
Phase-curve Parameters \\
\hline 
Average Relative Planetary Flux  & $f_{p}$ & ppm & $204\pm27$ \\
Planetary Phase-curve Amplitude & $A_{\mathrm{atm}}$ & ppm & $160.4\pm6.7$ \\
Planetary Phase-curve Offset & $\delta$ & deg. & $-0.7\pm2.2$ \\
Stellar Ellipsoidal Distortion Amplitude & $A_{\mathrm{ellip}}$ & ppm & $31.3\pm6.0$ \\
Secondary Eclipse Depth$^{\dag}$ & $D_{d}$ & ppm & $364\pm28$ \\
Nightside Flux$^{\dag}$ & $D_{n}$ & ppm & $43\pm27$ \\
  Dayside Brightness Temperature$^{\dag}$ & $T_{d}$ & K   & $3237\pm59$ \\
  Nightside Brightness Temperature$^{\dag}$ & $T_{n}$  & K     &  $1700^{+700}_{-1200}$\\
\hline
\hline 
Stellar Parameters  \\
\hline
  Effective Temperature   & $T_{\rm eff}$    & K     & $7300\pm100$ \\ 
  Metallicity       & [Fe/H]           & ---     & {$-0.1\pm0.12$} \\ 
  Surface Gravity & $\log g$ & --- & $4.1\pm0.2$ \\
  Projected Rotational Speed  & $v \sin i$       & \kms  & {$85.1\pm6.3$}          \\
  Stellar Mass              & $M_\star$         & \msun & $1.79 \pm 0.26  $ \\
  Stellar Radius            & $R_\star$         & \rsun & $1.950 \pm 0.048$ \\
\hline
\hline   
RV Parameters  \\
\hline
  RV Semiamplitude    & $K_s$          & \ms     & $<281 \, {(2\sigma)}$   \\
  Systemic Velocity & $V_{\rm sys}$  & \kms    & $-13.94\pm0.17$      \\
  \hline
\hline 
Planetary Parameters  \\
\hline
Planet Mass & $M_p$ & \mjup & $<2.3 \, {(2\sigma)}$ \\
Planet Radius & $R_p$ & \rjup &  $1.875\pm0.053$\\
Orbital Semimajor Axis & $a$ & au & $0.0389\pm0.0011$\\
Equilibrium Temperature & $T_{\rm eq}$ & K & $2492\pm38$\\
\hline
\end{tabular}
  \caption{Parameters for the TOI-1518 (TIC 427761355) planetary system. Relevant observing information is obtained from the TESS Quick Look Pipeline (QLP) and Atlas parameters. The V-band magnitude is obtained from the TESS input catalog \citep{Stassun2018}. The transit and phase curve parameters are simultaneously obtained from a joint fit of the full-orbit TESS light curve and ground-based full-transit photometric datasets (Section~\ref{subsec:joint}). Derived parameters (i.e., quantities not directly fit for in the light-curve analysis) are indicated by the superscript $^{\dag}$. The stellar parameters are determined by fitting a co-added high-resolution spectrum with a stellar model using \textsf{Spectroscopy Made Easy} and by a model fit to the broadband SED (Section~\ref{subsec:specmod}). The RV parameters are measured from FIES radial velocities  (Section~\ref{subsec:fies}).
  }
  \label{tab:par}
\end{table*}

\begin{figure}
    \centering
    \includegraphics[width=\linewidth]{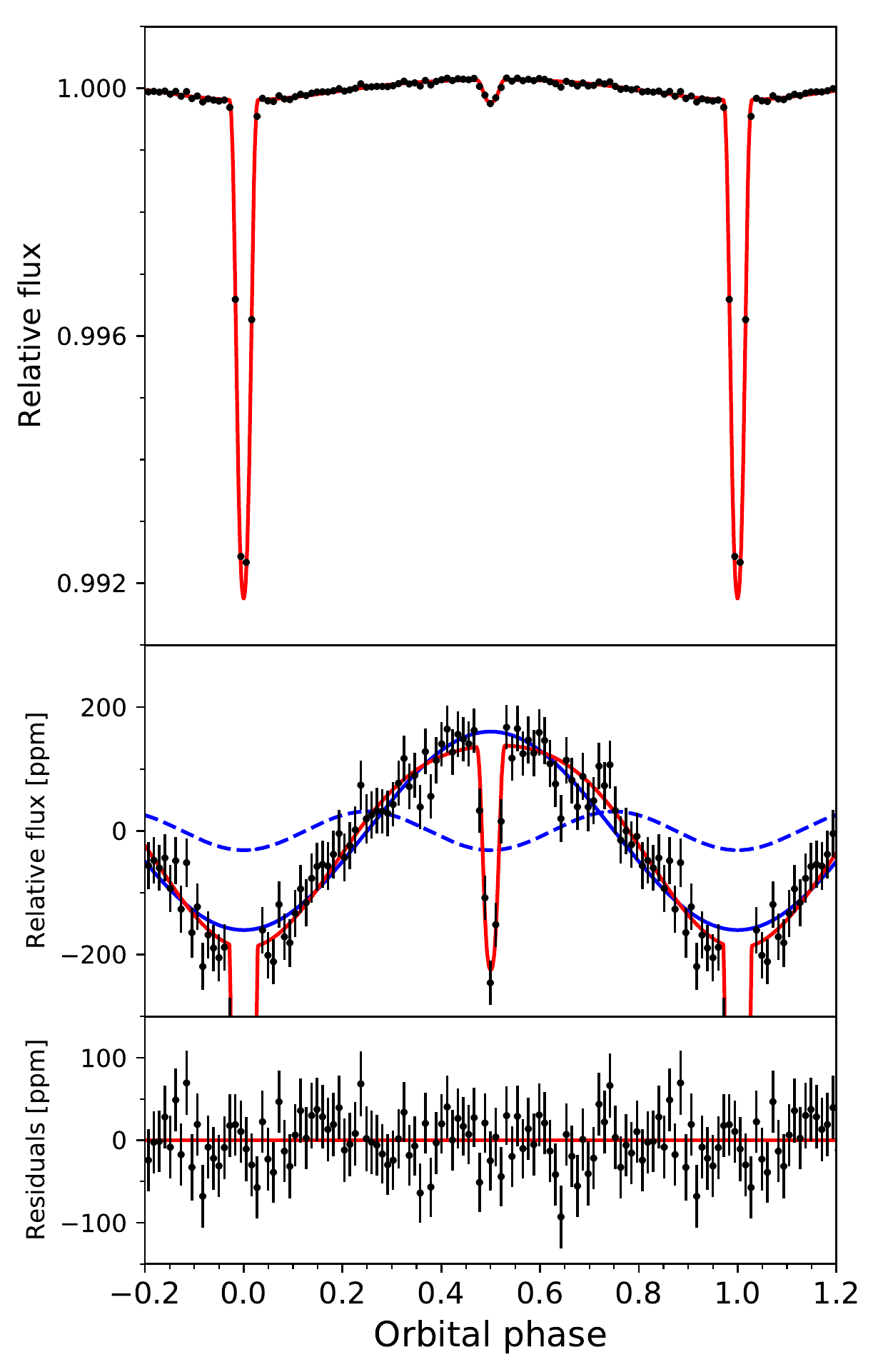}
    \caption{Top panel: systematics-corrected and phase-folded TESS light curve of TOI-1518, binned in 30-minute intervals, with the best-fit phase-curve model plotted in red. Middle panel: zoomed-in view of the phase-curve modulations and secondary eclipse. The atmospheric brightness modulation and ellipsoidal distortion signals are plotted separately in the solid and dashed blue lines. Bottom panel: corresponding residuals from the best-fit model.}
    \label{fig:tessfit}
\end{figure}

\begin{figure}
    \centering
    \includegraphics[width=\linewidth]{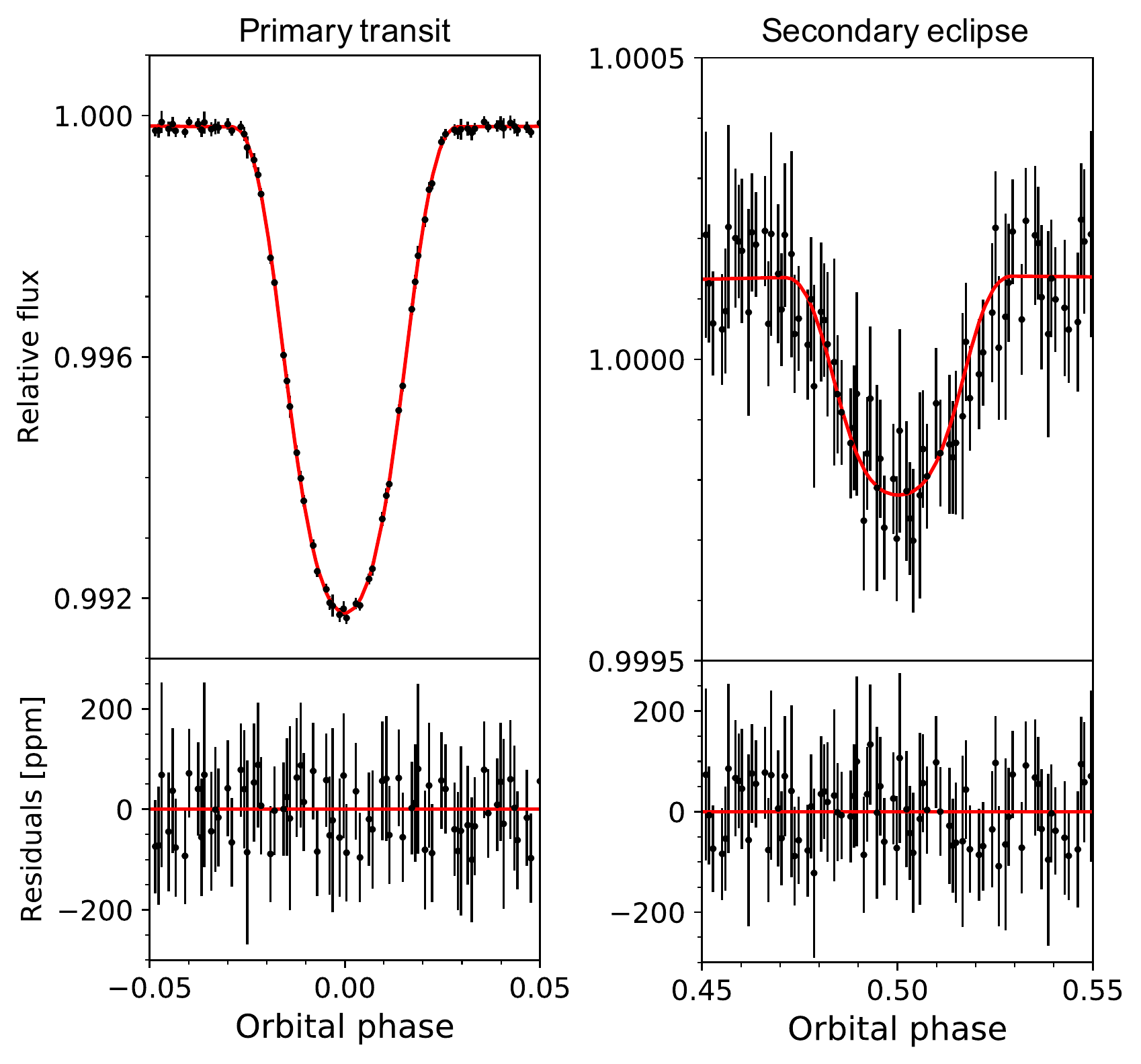}
    \caption{Zoomed-in views of the primary transit (left) and secondary eclipse (right) of TOI-1518b. The light curves are binned in 3-minute intervals. Note the difference in vertical scale between the two plots. The difference in out-of-occultation baselines primarily reflects the planetary phase-curve modulation. The bottom panels show the corresponding residuals from the best-fit model.}
    \label{fig:occult}
\end{figure}

The results of our joint fit are listed in Table~\ref{tab:par}. Figure~\ref{fig:tessfit} shows the binned, phase-folded, and systematics-corrected TESS light curve alongside the best-fit phase-curve model. Close-up views of the primary transit and secondary eclipse portions of the light curve are provided in Figure~\ref{fig:occult}. The secondary eclipse and phase-curve modulations are clearly discernible. {The detrended ground-based transit light curves are plotted in the Appendix.} 

{The orbital period of $1.902603\pm0.000011$ is measured to $\sim$1~s precision. We obtain a planet--star radius ratio of $R_{p}/R_\star=0.0988^{+0.0015}_{-0.0012}$, which is marginally more precise than the value derived from the TESS light curve alone. Likewise, we find slightly-improved values for the impact parameter and scaled semimajor axis: $b=0.9036^{+0.0061}_{-0.0053}$, $a/R_\star=4.291^{+0.057}_{-0.061}$. The secondary eclipse depth is measured to more than $12\sigma$ significance: $364\pm28$~ppm. The atmospheric phase-curve modulation has a semiamplitude of $160.4\pm6.7$~ppm. No significant phase shift in the planet's phase curve is measured, indicating that the location of maximum brightness on the dayside hemisphere is well-aligned with the substellar point. The derived nightside flux is $43\pm27$~ppm. The ellipsoidal distortion signal from the host star is detected at $5.2\sigma$ significance, with a semiamplitude of $31.3\pm6.0$~ppm. All of the phase-curve parameters are statistically identical to the values that we obtain from fitting the TESS light curve independently.} The planet's atmospheric brightness modulation and the star's ellipsoidal distortion signal are plotted separately in the middle panel of Figure~\ref{fig:tessfit}.

The full set of marginalized two-parameter posteriors for the fitted astrophysical quantities {(excluding the limb-darkening coefficients)} is plotted in the Appendix. As expected, due to the grazing nature of the transits and secondary eclipses, there are significant correlations between $b$, $R_p/R_\star$, and $f_p$, in addition to the typical degeneracy between $b$ and $a/R_\star$.

\subsection{SPP Speckle Interferometry}
\label{subsec:speckle}

\begin{figure}
\includegraphics[width=\linewidth, trim={1.3cm 0.3cm 1.3cm 0.3cm},clip]{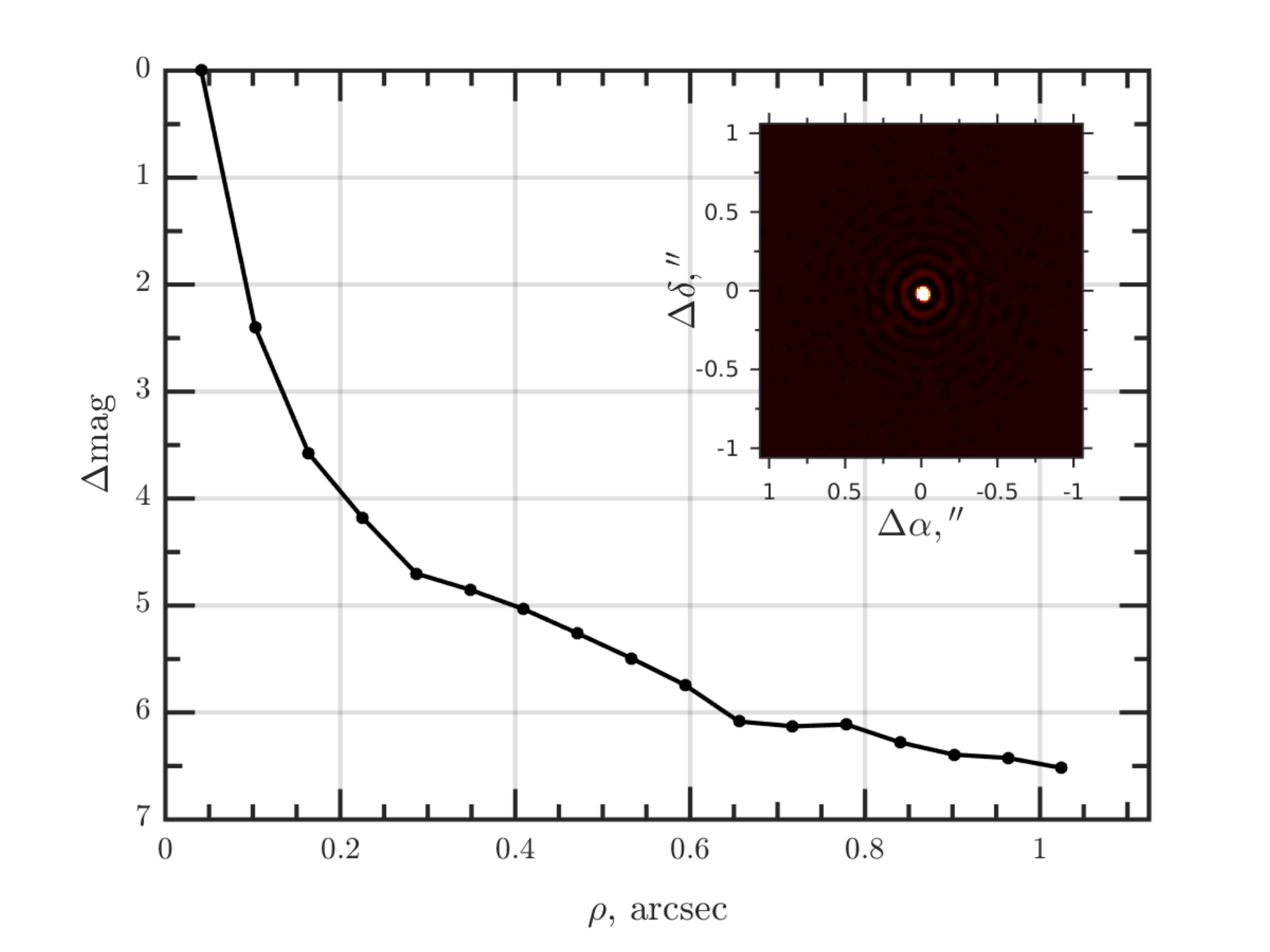}
\caption{SPP $5\sigma$ contrast curve for TOI-1518 with autocorrelation function (ACF) inset. The observations were obtained at $\lambda_\mathrm{c}=880$~nm (FWHM$=$70~nm). 
\label{fig:speckle_contrast}}
\end{figure}

TOI-1518 was observed using speckle interferometry on 2020 October 26 with the SPeckle Polarimeter (SPP; \citealt{Safonov2017}) on the 2.5\,m telescope at the Sternberg Astronomical Institute of Lomonosov Moscow State University (SAI MSU). The spectral band has a central wavelength of $880$~nm and a FWHM of 70~nm. The detector has a pixel scale of 20.6 mas px$^{-1}$, and the angular resolution was 89~mas. The detection limit for faint stellar companions is provided in Figure~\ref{fig:speckle_contrast}. We did not detect any companion brighter than this limit, e.g., 6.5 mag at $1^{\prime\prime}$.

\subsection{{EXPRES} Spectroscopy}
\label{subsec:expres}

EXPRES is an ultra-stable optical spectrograph recently commissioned at the Lowell Discovery Telescope \citep{Levine2012}. It is designed for extreme-precision radial velocity surveys \citep[see][for details about the instrument specifications and reduction pipeline]{Jurgenson2016, Blackman2020, Petersburg2020, Brewer2020} and also has the capacity for atmospheric characterization (see, for example, the recent study of ultra-hot Jupiter MASCARA-2b by \citealt{Hoeijmakers2020}). One transit of TOI-1518b was observed on the night of 2020 August 2, involving 41 $\sim$300~s exposures. The extracted spectra have a signal-to-noise (S/N) of $\sim$20--40 for pixels in the continuum. Orders were {continuum normalized \citep{Petersburg2020}, and subsequently} stitched together to form one-dimensional spectra. Telluric absorption from O$_2$ and H$_2$O in Earth's atmosphere was corrected with \textsf{molecfit} \citep{smette2015} in the geocentric rest-frame using similar fitting parameters as \citet{Allart2017}. {Indeed, telluric modeling with \textsf{molecfit} has become a frequent step in high-resolution optical atmosphere studies \citep[e.g.][]{CasasayasBarris2019}, and is advantageous over empirical models for resolving some atmospheric spectral features \citep{Langeveld2021}}.

\subsection{Spectroscopic Modeling}
\label{subsec:specmod}

Before analyzing the transit, we used \textsf{Spectroscopy Made Easy} (SME {423}; \citealt{Valenti1996}) to infer stellar parameters from the high-resolution spectra. {The analysis closely follows that of \citet{Brewer2016}, including the choice of fitting parameters and wavelength segments.} The model made use of a VALD3 line-list \citep{Ryabchikova2015}, an \textsf{ATLAS9} atmospheric model \citep{Kurucz1993, Heiter2002}, and a Gaussian convolution instrument profile with $R = 137,000$. {Microturbulence was fixed at 0.85 \kms, and macroturbulence was scaled to $T_{\rm eff}$ following the parametrization of \citet{Brewer2016}. However, the fit was largely insensitive to these parameters since the broadening is completely dominated by stellar rotation. The rotational broadening also prevents a robust fit to abundances of individual species. We opt to solve for a global [M/H] with the assumption of a solar abundance pattern for individual elements.}

The true uncertainties on effective temperature ($T_{\rm eff}$), metallicity ([Fe/H]), and rotation speed ($v\sin i$) are difficult to gauge \citep{Piskunov2017}. The Levenberg-Marquardt optimization algorithm involves computing a curvature matrix at the minimum of the objective function, the inverse of which is the covariance matrix. The square root of the diagonal elements are the formal uncertainties on the parameters, assuming that the dominant source of uncertainty is from measurement errors (i.e. Poisson statistics on the spectrum). The actual uncertainty is dominated by systematic effects and model errors, as opposed to measurement errors. \citet{Piskunov2017} describe a method to incorporate model errors. It involves measuring the sensitivity of each spectral pixel to changes in the parameters and estimating the change necessary to reduce the fit residuals to zero. The cumulative distribution function (CDF) of these parameter perturbations is then calculated. The central region of each CDF gives an estimate of the model error. \citet{Piskunov2017} discuss this method in greater detail, and we adopt it for our analysis. 

We find that TOI-1518 is a rapidly rotating F0 star with $v\sin i = 85\pm6$ \kms, which agrees with expectations for this spectral type \citep{Nielsen2013}. A fitted [Fe/H] of $-0.1 \pm 0.12$ is low for a star hosting a hot Jupiter \citep{Fischer2005}; only $\sim 4\%$ of planet hosts have [Fe/H] near $-0.1$. However, the uncertainties on [Fe/H] are large due to the widening and blurring of spectral lines (a consequence of the rapid rotation), so the star may be more metal-rich than the best-fit value suggests. The best fit effective temperature {and surface gravity are $T_{\rm eff} = 6910\pm445$ K and $\log g = 3.97\pm0.62$, respectively}. More detailed investigation of the stellar spectrum might warrant modeling non-LTE effects in the deepest lines and calibrating line positions and $\log gf$ values. However, these considerations are most important for cooler stars with total rotational broadening $\lesssim 10$ \kms\ \citep{Brewer2016}, and their impact on TOI-1518 is reduced due to the rotation speed. 
{Measurements of $v\sin i$ and [Fe/H] are listed in Table~\ref{tab:par}. However, we opt to report the better constrained measurements of $\log g$ and $T_{\rm eff}$} from our spectral energy distribution modeling (see below). Our inferred $v\sin i$ is used to analyze the Rossiter-McLaughlin (RM) effect \citep{Rossiter1924, McLaughlin1924} in Section~\ref{subsec:rm}. 

\begin{figure}[!t]
\includegraphics[width=\linewidth,trim=85 70 85 90,clip]{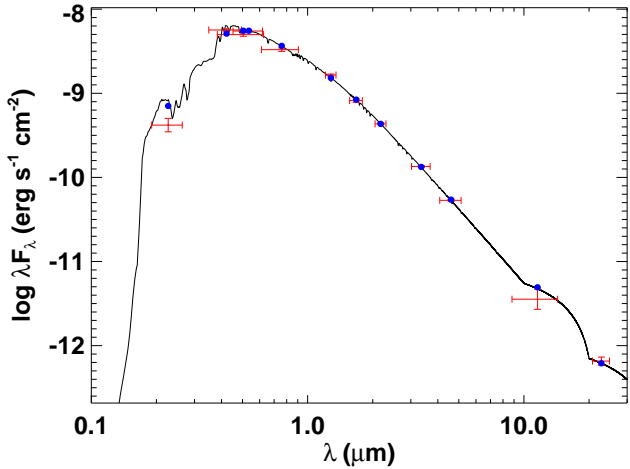}
\caption{Spectral energy distribution of TOI-1518. Red symbols represent the observed photometric measurements, where the horizontal bars represent the effective width of the passband. Blue symbols are the model fluxes from the best-fit Kurucz atmosphere model (black).  \label{fig:sed}}
\end{figure}

As an independent determination of the stellar parameters, we performed an analysis of the broadband spectral energy distribution (SED) of the star together with the {\it Gaia\/} DR2 parallaxes \citep[adjusted by $+0.08$~mas to account for the systematic offset reported by][]{StassunTorres:2018}, following the procedures described in \citet{Stassun:2016,Stassun:2017,Stassun:2018}. We took the $B_T V_T$ magnitudes from {\it Tycho-2}, the $BVi$ magnitudes from {\it APASS}, the $JHK_S$ magnitudes from {\it 2MASS}, the W1--W4 magnitudes from {\it WISE}, the $G G_{\rm BP} G_{\rm RP}$ magnitudes from {\it Gaia}, and the NUV magnitude from {\it GALEX}. Together, the available photometry spans the full stellar SED over the wavelength range 0.2--22~$\mu$m (see Figure~\ref{fig:sed}).  

We performed a fit using Kurucz stellar atmosphere models, with the free parameters being the effective temperature ($T_{\rm eff}$), metallicity ([Fe/H]), surface gravity ($\log g$), and extinction ($A_V$); the extinction was restricted to the maximum line-of-sight value from the dust maps of \citet{Schlegel:1998}. The resulting fit (Figure~\ref{fig:sed}) has a $\chi^2$ of 20.3 (with 12 degrees of freedom) and best-fit parameters $T_{\rm eff} = 7300\pm100$~K, $\mathrm{[Fe/H]} = 0.0\pm0.2$, $\log g = 4.1\pm0.2$, and $A_V = 0.05 \pm 0.05$. The relatively low $A_V$ may be surprising considering the low galactic latitude; however, this $A_V$ is consistent with the 3D dust maps for this system's position from \citet{Green:2019}.

Integrating the (unreddened) model SED gives the bolometric flux at Earth, $F_{\rm bol} = (6.52 \pm 0.31) \times 10^{-9}$ erg~s$^{-1}$~cm$^{-2}$. Taking $F_{\rm bol}$ and $T_{\rm eff}$ together with the {\it Gaia\/} DR2 parallax ($4.398 \pm 0.033$ mas) gives a stellar radius of $R_\star = 1.950 \pm 0.048$~R$_\odot$. In addition, we use $R_\star$ together with $\log g$ to obtain an empirical mass estimate of $M_\star = 1.79 \pm 0.26$~M$_\odot$, which is consistent with that calculated via the empirical relations of \citet{Torres:2010} --- $M_\star = 1.70 \pm 0.12$~M$_\odot$. 

\subsection{FIES Spectroscopy}
\label{subsec:fies}
Starting on June 14th 2020 and ending on February 3rd 2021, we monitored TOI-1518 with the Nordic Optical Telescope \citep[NOT;][]{Djupvik2010} using the FIber-fed Echelle Spectrograph (FIES; \citealt{Telting2014}). This was done in order to constrain the out-of-transit Doppler motion of the star, although the high rotation rate of the star broadens the spectral lines and makes it difficult to measure. The FIES high-resolution fiber reaches $R \sim 67,000$ and covers wavelengths from 3760\,\AA{} to 8840\,\AA{} with no gaps below 8200\,\AA. We obtained 22 spectra, which we extract as described in \citet{Buchhave2010} and assign wavelengths using ThAr calibrations taken immediately before and after each exposure. The SNR per resolution element ranges from 49 to 141, measured in the 5500\,{\AA} spectral order. {We did not include RVs from the EXPRES spectra when constraining the Doppler motion, as this would require an extra instrumental offset parameter for a single night of data.}

To extract the radial velocities from the FIES spectra, we perform a least-squares deconvolution (LSD) analysis to derive the spectroscopic broadening profiles from each observation \citep{Donati1997}. We deconvolve each spectrum against a synthetic non rotating spectral template generated via the ATLAS9 library \citep{Castelli2003}, and fit the resulting line profiles with a kernel incorporating the rotational, instrumental, and macroturbulent components of the line broadening function, similar to the recent analysis of HAT-P-70 by \citet{Zhou2019a}. The extracted RVs are listed in Table~\ref{tab:rvfies}.

One point is excluded from the analysis, since it overlaps with the transit. Using the \textsf{radvel} package \citep{Fulton2018}, we model the orbit as circular with no other planets in the system; the stipulation of a circular orbit is in line with the results of our TESS light-curve fit, which indicated a $2\sigma$ upper limit on orbital eccentricity of 0.01 (Table~\ref{tab:par}). We define Gaussian priors for period and time of conjunction (using the values and uncertainties from Table~\ref{tab:par}), as well as a broad, uniform prior on the RV semiamplitude $K_s$. We sample the parameter space with an MCMC analysis using the default \textsf{radvel} setup and let the software run until it determines that the chains are well-mixed. 

The $K_s$ posterior distribution peaks near its median at 152{\,\ms} with a $1\sigma$ error of 75{\,\ms}, i.e. less than $2\sigma$ significance. We derive a 95\% upper limit of 281{\,\ms}. Adopting the stellar mass determined in Section~\ref{subsec:specmod} and the orbital inclination determined in Section~\ref{subsec:rm}, this corresponds to a planetary upper mass limit of 2.3{\,\mjup}, well within expectations for hot Jupiters.

\begin{figure}
    \centering
    \includegraphics[width=\linewidth]{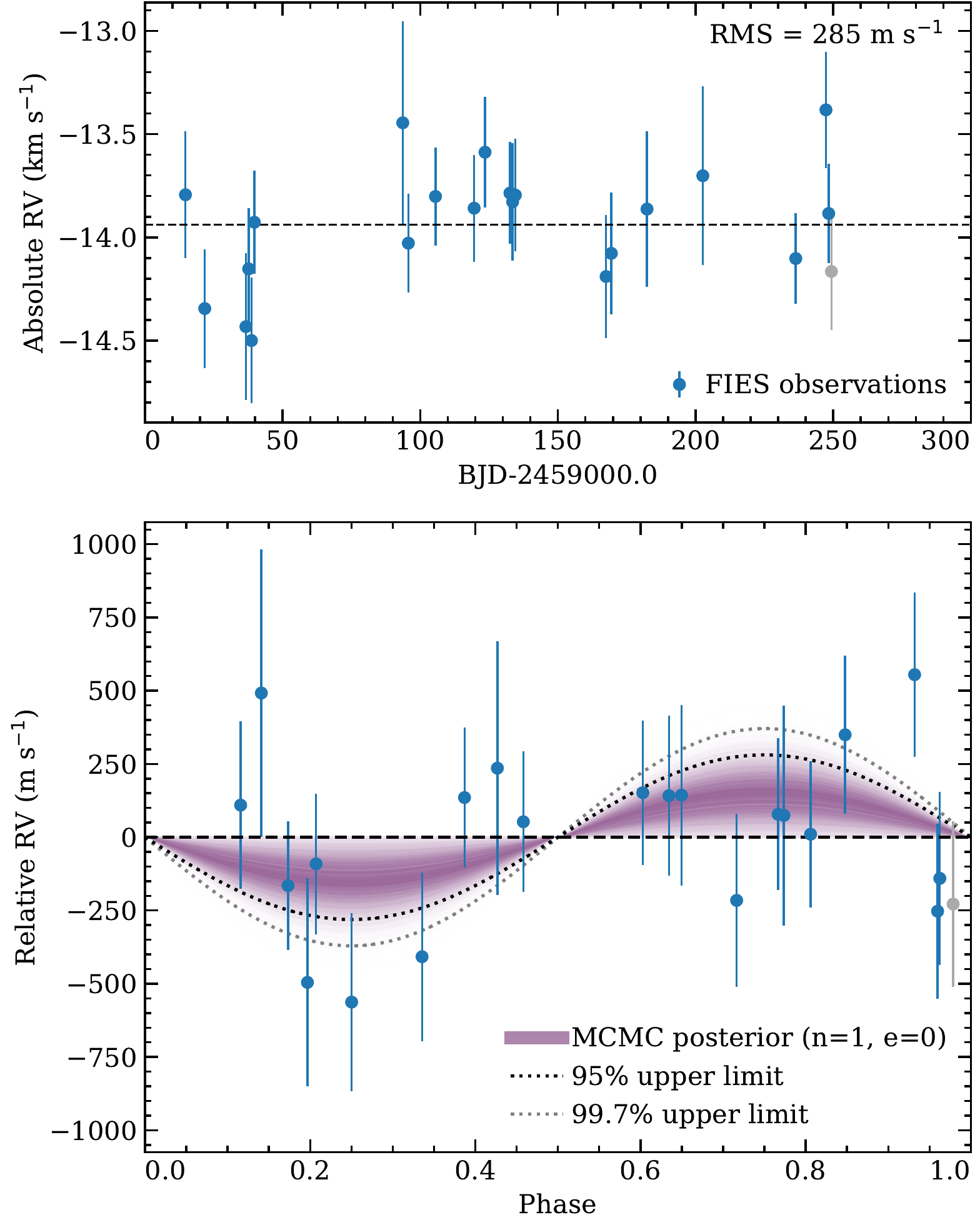}
    \caption{Out-of-transit RVs measured with FIES. Upper panel shows the full RV time series. Lower panel shows the same RVs phase-folded from $T_c$ with the known orbital period. The $K_s$ posterior distribution is visualized as shaded, purple curves in the background (darker: higher density). The last observation (gray) overlaps with the transit and has therefore been excluded from the fit. While the data have large uncertainties, the amplitude of the velocity variation is consistent with a planetary companion of $M_p<2.3~M_{\rm J}$.}
    \label{fig:rv}
\end{figure}

To determine the systemic velocity, we compute the weighted mean of the measured RVs, $-14.79 \pm 0.06\,\kms$, which must be corrected for an instrumental offset of $-0.87 \pm 0.16\,\kms$, found from standard stars. We arrive at a systemic velocity $V_\textrm{sys}$ of $-13.94 \pm 0.17\,\kms$. The derived RVs are displayed in Figure~\ref{fig:rv}, with the posterior distribution of $K_s$ visualized along with the phase-folded velocities. {The observations provide generally good sampling of the orbital phase, and have mean cadence of $11.2$ days between adjacent observations; we do not expect the RV signature to arise from sampling artifacts or aliases.} More data is needed {though} to determine if the scatter in the RVs could be caused by one or more additional planets in the system.

\setlength{\tabcolsep}{8.0pt}
\begin{table}
\begin{tabular}{c c c c}
\hline
Time (BJD) & Phase & $v$ (\kms) & $\sigma_{v}$ (\kms) \\
\hline
2459014.69572 & 0.65 & -14.64 & 0.31 \\
2459021.70842 & 4.34 & -15.19 & 0.29 \\
2459036.66525 & 12.20 & -15.28 & 0.36 \\
2459037.65405 & 12.72 & -15.00 & 0.30 \\
2459038.66970 & 13.25 & -15.34 & 0.30 \\
2459039.72700 & 13.81 & -14.77 & 0.25 \\
2459093.63742 & 42.14 & -14.29 & 0.49 \\
2459095.66605 & 43.21 & -14.87 & 0.24 \\
2459105.52107 & 48.39 & -14.64 & 0.24 \\
2459119.56182 & 55.77 & -14.70 & 0.26 \\
2459123.52136 & 57.85 & -14.43 & 0.27 \\
2459132.56838 & 62.60 & -14.63 & 0.25 \\
2459133.54446 & 63.12 & -14.67 & 0.29 \\
2459134.53120 & 63.63 & -14.64 & 0.27 \\
2459167.49441 & 80.96 & -15.03 & 0.30 \\
2459169.40210 & 81.96 & -14.92 & 0.29 \\
2459182.36116 & 88.77 & -14.71 & 0.38 \\
2459202.62976 & 99.43 & -14.54 & 0.43 \\
2459236.39450 & 117.17 & -14.95 & 0.22 \\
2459247.35102 & 122.93 & -14.23 & 0.28 \\
2459248.35228 & 123.46 & -14.73 & 0.24 \\
2459249.34239 & 123.98 & -15.01 & 0.28 \\
\hline
\end{tabular}
  \caption{Radial velocities of TOI-1518 extracted from FIES spectra. Columns correspond to the timestamp of the exposure, {orbital phase}, velocity, and uncertainty on velocity.}
  \label{tab:rvfies}
\end{table}

\section{The Spectroscopic Transit} \label{sec:cca}

In this section, we describe the methods used to analyze the spectroscopic transit observation from EXPRES. Cross-correlation was performed with the \textsf{X-COR} pipeline, previously used for atmospheric detections in WASP-121b \citep{Cabot2020, BenYami2020} and MASCARA-2b \citep{Hoeijmakers2020}. Cross-correlation has become a standard approach for exoplanet atmospheric analyses at high-resolution \citep[e.g.,][]{Snellen2010,Brogi2012,Birkby2013}. This method relies on resolving the orbital motion of the planet via its Doppler shift on absorption lines (or more recently emission lines, as shown by {\citealt{Nugroho2017}} and \citealt{Pino2020}). While individual lines are generally low-S/N, their contributions may be stacked by cross-correlating an atmospheric model with the data. Then, one can analyze the resultant cross-correlation function (CCF). This technique has led to a slew of molecular detections in the near-infrared (NIR), as well as atomic and ion detections in the optical, starting with KELT-9b \citep{Hoeijmakers2018}. Please see \citet{Madhusudhan2019} and \citet{BenYami2020} for more examples of recent atmospheric detections at high-resolution. We briefly discuss the relevant methods in the following subsection. We then turn our attention to the RM effect and atmospheric signals present in the CCFs.

\subsection{Detrending and Cross-Correlation}

The most prominent features in the time-series spectra of TOI-1518b are absorption lines originating in the stellar photosphere, as well as telluric lines caused by Earth's atmosphere. As mentioned above, we corrected tellurics by fitting and dividing each spectrum by a \textsf{molecfit} model. {The spectra were then linearly interpolated onto a common 0.01 \AA\ wavelength grid in the barycentric rest-frame.} {We observed a significant narrow sodium absorption component in the original spectra, which is likely due to the interstellar medium. Next, we co-added all out-of-transit spectra into a master $F_{\rm out}$ and then divided each individual spectrum by $F_{\rm out}$. Interstellar medium features were removed through division by $F_{\rm out}$ since we opted to not correct for the RV motion of the star \citep{Casasayas-Barris2018}. Since stellar lines are significantly broadened from rotation, the RV motion has negligible effect on the planet's transmission spectrum.} Remaining broadband variations in the spectra were removed by a high-pass Gaussian filter with a standard deviation of 75 pixels. We restricted our analysis to the region $4000-6800$ \AA. The S/N falls off at bluer wavelengths, and redder wavelengths suffer from particularly severe telluric absorption. Throughout the analysis, about {$1\%$ of the data were masked to avoid particularly low S/N pixels on the blue edge of the spectrum and within Balmer lines.}

Cross-correlation was performed between each transmission spectrum and a continuum-subtracted \textsf{PHOENIX} stellar model \citep{Husser2013}. The model parameters were selected from a grid and chosen to be close to the inferred parameters: $T_{\rm eff} = 7000$ K, $\log g = 4.0$ and [Fe/H] $= 0.0$. The CCF is essentially a sliding dot product between the observed spectra and the model template. It is defined as a function of time $t$ and velocity $v$:
\begin{equation}
    {\rm CCF}(v,t) = \frac{\sum_i{f(i|t) m(i|v)w(i)}}{\sum_i m(i|v)w(i)}.
    \label{eqn:cc}
\end{equation}
Here, the observed spectrum $f(i|t)$ corresponds to the flux in pixel $i$ at time $t$. The \textsf{PHOENIX} stellar template, denoted by $m(i|v)$, has been Doppler shifted by some velocity $v$ and is interpolated onto the observed wavelength grid. The weighting term $w(i)$ is chosen to be the inverse time variance of each pixel, so as to downweight contributions from pixels previously in the cores of stellar or telluric lines. The CCF velocities are a grid spanning $-500$ to $+500$ \kms\ in increments of 2 \kms. 

\subsection{Spin-Orbit Misalignment}
\label{subsec:rm}

\begin{figure}
    \centering
    \includegraphics[width=\linewidth]{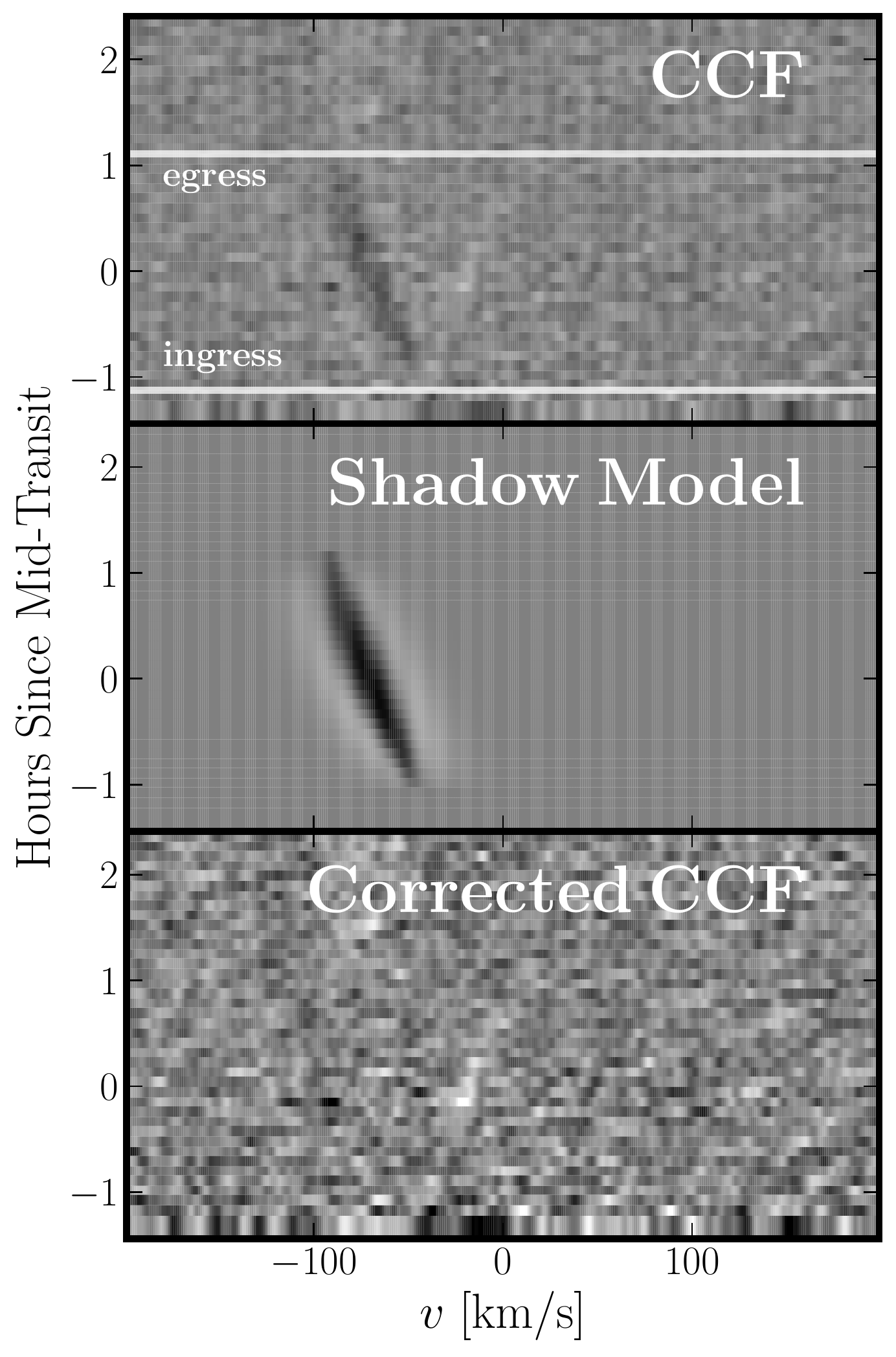}
    \caption{Cross-correlation function between the \textsf{PHOENIX} stellar template and individual transmission spectra. {\it Top Panel}: CCF annotated with the start of ingress and egress. The Doppler shadow (dark) and atmospheric trail (light) form a ``V" shape with a vertex at about $-30$ \kms. {\it Middle Panel}: Doppler shadow model as described in the text. {\it Bottom Panel}: Corrected CCF where the Doppler shadow model has been subtracted.}
    \label{fig:ccf}
\end{figure}

Although we have isolated the planetary atmospheric transmission spectrum, there are residuals at former locations of stellar lines that arise from the division by $F_{\rm out}$. While $F_{\rm out}$ is a good template for the out-of-transit stellar spectrum, the stellar line profiles during transit are distorted because the planet occults part of stellar disk. The projected location of the planet against the stellar disk changes throughout the transit, dependent on its impact parameter $b$ and projected obliquity $\lambda$. The star has a projected rotation speed $v\sin i$, and the flux emitted at each point on the star's surface is Doppler shifted by some local velocity. The transit removes part of the integrated stellar flux, and breaks the symmetry between each side of the rotating star. This phenomenon is known as the Rossiter-McLaughlin effect. It is observed by the apparent ``Doppler shadow'' in the CCFs \citep{Collier-Cameron2010a}, where a dark trail traces the local velocity of the occulted stellar region. 

We model the shadow in a similar fashion as \citet{Hoeijmakers2020} and show the steps in Figure~\ref{fig:ccf}. First, we fit a {double-Gaussian} profile {(sum of two Gaussians)} to the Doppler shadow in each CCF row and record the {inner profile's} fitted mean, standard deviation, and amplitude. {The inner profile models the core of the Doppler shadow, whereas the outer profile models positive wings on either side that result from normalizing the spectra. The second Gaussian's mean was fixed to that of the first, and the standard deviation was fixed to 18 \kms.} {A third degree polynomial} is then fit to the means as a function of time, and then evaluated at the times of each exposure. This step was repeated for the {remaining fitted} parameters. Finally, the Doppler shadow was modeled as a series of {double-}Gaussian profiles, with parameters determined by the above polynomials. The polynomials ensure that the model smoothly varies in time. While this is not a sophisticated physical model of the shadow, it is effective at correcting the CCF so that the Doppler shadow does not adversely affect the atmospheric analysis. Serendipitously, the shadow and planetary signal do not overlap except for a small window at the start of transit. This configuration is only possible when the planet's path is roughly parallel to the projected stellar rotation axis and the transit takes place near the limb of the star. Nevertheless, it is still important to model out the Doppler shadow to correctly interpret the S/N of the atmospheric signal.

The path traced out by the Doppler shadow provides additional constraints on the transit geometry \citep{Collier-Cameron2010, Bourrier2015, Cegla2016}. The portion of the stellar disk occulted by the planet has a local velocity
\begin{equation}
    v_\star(t) = x_{\perp}(t) v \sin i. 
\end{equation}
The orthogonal distance $x_{\perp}$ is determined by the position of the planet:
\begin{equation}
    x_{\perp}(t) = x_p(t) \cos (\lambda) - y_p(t) \sin (\lambda)
\end{equation}
\begin{equation}
    x_p(t) = \frac{a}{R_\star} \sin (2 \pi \phi)
\end{equation}
\begin{equation}
   y_p(t) = -\frac{a}{R_\star} \cos (2 \pi \phi) \cos (i_p).
\label{eqn:yp}
\end{equation}
Therefore, we can obtain independent constraints on ${a}/{R_\star}$, $\lambda$, $v \sin i$, and $i_p$ from the light curve and spectrum fitting (note the distinction between $i_p$ and stellar inclination $i$, the latter of which we do not investigate here; however it also may be probed by considering differential rotation \citep{Cegla2016}). We run an MCMC routine that samples these parameters and fits the path of the shadow described by the polynomial fit described above. As an initial check, we use uniform priors: $2 < {a}/{R_\star} < 12$, $0 < \lambda < 2\pi$ and $0 < i_p < \pi$. We define Gaussian priors for the rotation speed and global offset: $v \sin i \sim \mathcal{N}(\mu=80,\: \sigma=50)$ \kms, $V_{\rm sys} \sim \mathcal{N}(\mu=-14.5,\: \sigma=2)$ \kms. The results are not strongly dependent on the choice of prior for the global offset, owing mainly to the large rotation speed.
The sampler includes 15 walkers with 50,000 steps each. We set the uncertainty on each point equal to the standard deviation of the Gaussian profile. We assume that the difference between each data point and the model is independent and normally distributed. We discard the first 5,000 steps and thin the chains by a factor of 40 (approximately the autocorrelation time). 

From this initial analysis, we obtain a scaled semimajor axis $a/R_\star = {2.95^{+0.95}_{-0.72}}$. The inclination is in better agreement with Table~\ref{tab:par}, at $i_p = {76.1^{+3.3}_{-4.9}}$ degrees. We also note a strong correlation between $\lambda$ and $i_p$. Next, we rerun the MCMC using photometrically-derived priors on $a/R_\star$ and $i_p$ in order to establish a tighter constraint on obliquity. The final results of our MCMC analysis, listed in Table~\ref{tab:rmpar}, show that TOI-1518b is a highly-misaligned, retrograde planet, with $\lambda = {240.34^{+0.93}_{-0.98}}$ degrees. Indeed, close-in gas giants around hot stars are commonly misaligned \citep{Winn2010}. Companions with mass $\gtrsim 3$ \mjup\ around hot stars are less likely to be found in retrograde orbits \citep{Hebrard2011, Triaud2018}, but the RV-derived mass of TOI-1518b is below this threshold. 

\setlength{\tabcolsep}{6.0pt}
\begin{table}
\begin{tabular}{l  | c c r }
\hline
RM Parameter & Symbol & Units & Value \\
\hline 
  Scaled Semimajor Axis    & $a/R_\star$& -        & ${{4.272^{+0.058}_{-0.057}}}$ \\
  Proj. Obliquity          & $\lambda$	& deg.     & ${240.34^{+0.93}_{-0.98}}$    \\
  Orbital Inclination      & $i_p$	    & deg.     & ${77.92\pm0.24}$              \\
  Proj. Rot. Speed         & $v\sin i$	& \kms     & ${74.4\pm2.3}$                \\
\hline
\end{tabular}
  \caption{Rossiter McLaughlin (RM) parameters, inferred by fitting the path traced by the Doppler shadow in Section~\ref{subsec:rm}. We used the physical model of \citet{Cegla2016} and the \textsf{emcee} sampler \citep{Foreman-Mackey2013}. Free parameters included the above four as well as $V_{\rm sys}$, which returned a posterior distribution that was very similar to its prior Gaussian distribution. The parameters $a/R_\star$ and $i_p$ were constrained by Gaussian priors derived from the results of our TESS light-curve fit (Table~\ref{tab:par}).}
  \label{tab:rmpar}
\end{table}

\subsection{$K_p-V_{\rm sys}$ Analysis}

Closer inspection of Figure~\ref{fig:ccf} shows a faint, white trail spanning approximately $\pm 50$ \kms. This feature is a signature of the planet's atmosphere. Throughout the transit, the planet's apparent radial velocity changes as it moves towards and then away from the observer, given by
\begin{equation}
   v_p(t) = -K_p\sin(2\pi (t - T_c)/P),
   \label{eqn:vp}
\end{equation}
where $K_p$ is the semiamplitude of the {planet's radial velocity}. Because the planet orbits close in, the change in velocity is of order tens of \kms. The CCF at each time $t$ peaks when the \textsf{PHOENIX} model template is Doppler shifted by the planet's velocity, and features in the model line up with features in the actual transmission spectrum. The result is a trail in the CCFs that traces out a small portion of a sinusoidal curve. The planetary signal may be further enhanced by aligning and co-adding CCF rows, thus stacking the peaks and improving the signal's S/N. The slope of the CCF trail near transit is completely determined by $K_p$ through Equation~\ref{eqn:vp}. It is also offset from 0 by the systemic velocity $V_{\rm sys}$. It is useful to determine $K_p$ and $V_{\rm sys}$ by sampling values from a grid and attempting to shift and stack the CCFs for each combination of values \citep{Brogi2012}. The signal is maximized at the correct set of values.

The CCF trail only appears if the cross-correlation template contains features present in the planet's transmission spectrum. The trail in Figure~\ref{fig:ccf} indicates that the atmosphere contains neutral and/or ionized species present in the \textsf{PHOENIX} spectrum. The absorption line positions and relative strengths are unique to each species. Therefore, we can cross-correlate with a model template containing only one species, and then perform the $K_p-V_{\rm sys}$ analysis to search for an atmospheric signal. If the stacked CCF contains a sufficiently high significance peak, then we confirm the presence of that species in the atmosphere of the planet. Here, we define detection significance (S/N) as the number of standard deviations that the CCF peak lies away from mean of all values, for all combinations of $K_p$ and $V_{\rm sys}$. Many species of interest are present in the stellar spectrum and have a Doppler shadow in their CCFs. Therefore, after cross-correlating with each model template, we scale the shadow model obtained in Section~\ref{subsec:rm} by a best-fitting constant value and subtract it from the the CCF.

\subsection{Transmission Spectrum Model}

{During a planet transit, a fraction of the stellar light is filtered by the planetary atmosphere. To compute the high-resolution transmission spectra of the planet’s atmosphere, we first need to calculate the opacities of the elements in the atmosphere. In this work, the Fe and Fe$^+$ opacities were computed using the HELIOS-K software (\citealt{2021Grimm}). Our models for Fe and Fe$^+$ make use of the line-list tables from \cite{2018Kurucz}. The lines for both Fe and Fe$^+$ were computed assuming Voigt profiles, 0.032 cm$^{-1}$ spectral resolution, and a fixed line cutoff of 100 cm$^{-1}$. To calculate the transmission spectra, we developed our code based on the simple formalism presented in \cite{2017Gaidos} and \cite{2019Bower}. Our model computes the effective tangent height in an atmosphere that was discretised in 200 annuli. The model included some simplifications due to the unknown composition of the atmosphere of TOI-1518b and a weakly constrained planet bulk density: we assumed a surface gravity of $\log g = 3$ and an atmosphere in chemical equilibrium. The chemical calculations were done with the open-source code FastChem (\citealt{2018Stock}), assuming solar metallicities. We include in our model the H$^{-}$ bound–free and free–free absorption from \cite{1988John}. As shown in \cite{Kitzmann2018}, the H$^{-}$ continuum in UHJs is generally between 1 mbar and 10 mbar. Each high-resolution transmission spectrum includes Fe or Fe$^{+}$ along with H$^{-}$ continuum absorption and scattering by H and H$_2$. We generated a grid of high-resolution transmission spectra assuming isothermal atmospheres ranging from 2000 to 4000 K in steps of 500 K. Following subtraction of the continuum { with a sliding maximum filter and convolution with a Gaussian filter to match the EXPRES instrumental resolution}, these models serve as cross-correlation templates.}

\section{Atmospheric Characterization} \label{sec:atm}

\subsection{Detections}

\begin{figure*}
    \centering
    \includegraphics[width=\linewidth]{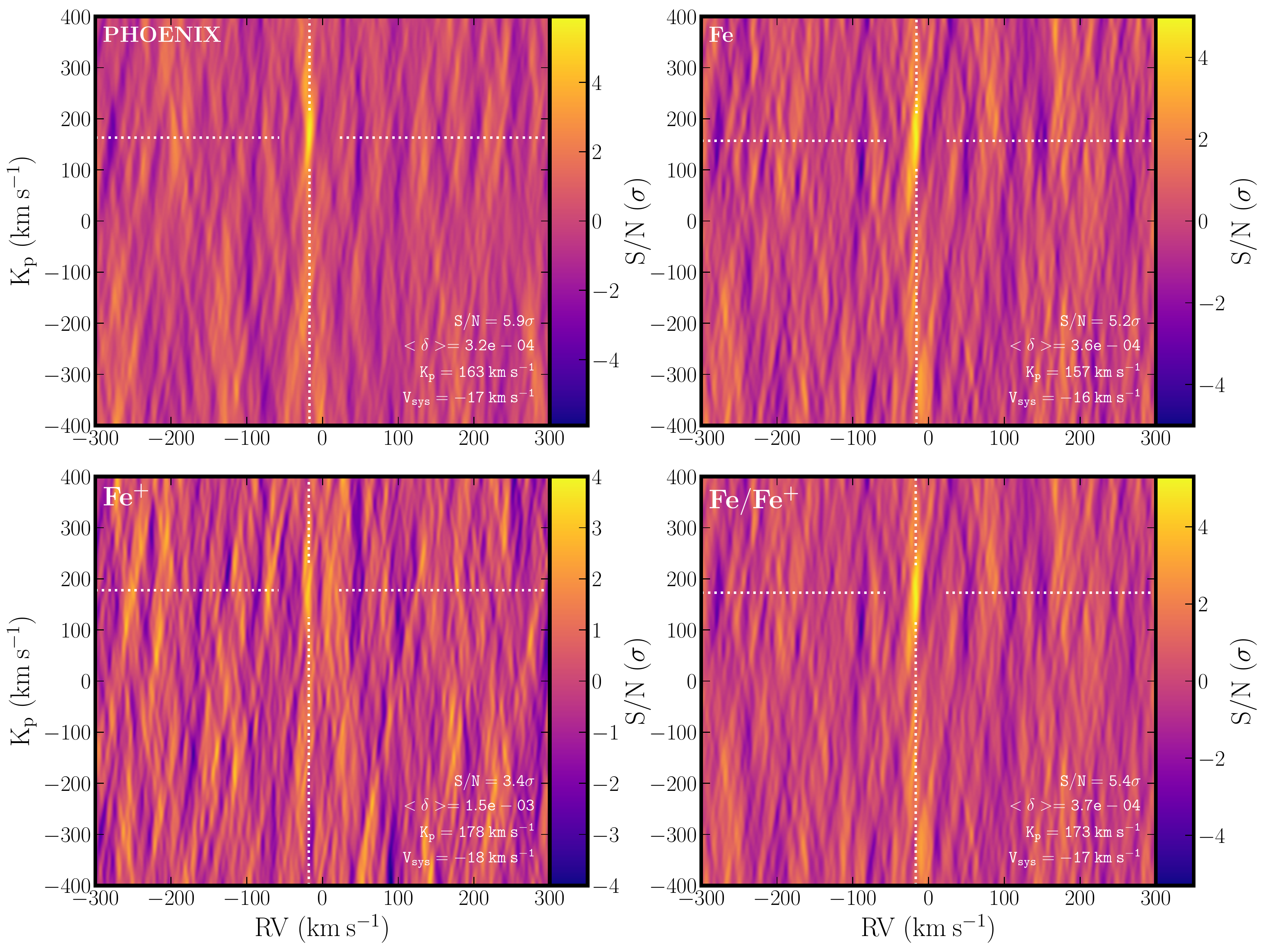}
    \caption{Atmospheric detections in TOI-1518b and their $K_p-V_{\rm sys}$ maps. The top-left corner {of each panel} indicates the cross-correlation template, and the bottom right corner lists properties of the peak value, including S/N of the detection, the average absorption depth after co-adding CCF rows ($\langle\delta\rangle$), and the maximal value of $K_p$, and $V_{\rm sys}$. {In all panels the Doppler shadow has been corrected per the methods in Section~\ref{subsec:rm}. For reference, we show results from cross-correlation with the \textsf{PHOENIX} spectrum used to model the Doppler shadow, revealing there are species common to both the planet's atmosphere and star}. The subsequent panels show results from cross-correlating with templates containing Fe and Fe$^+$. In each panel, the white dotted lines indicate the $K_p$ and $V_{\rm sys}$ with the highest signal.}
    \label{fig:det}
\end{figure*}

We detect Fe in the atmosphere of TOI-1518b at the ${5.2}\sigma$ level. We also report evidence of Fe$^+$ at the ${3.4}\sigma$ level. The \textsf{PHOENIX} model, which contains both species in addition to other atoms and ions, yields an enhanced atmospheric detection at $5.9\sigma$ confidence, {while a combined Fe/Fe$^+$ model yields a ${5.4}\sigma$ detection}. The Doppler shadow correction removes an artifact that otherwise biases detection significances. The $K_p$ and $V_{\rm sys}$ corresponding to the peak value are consistent across the various templates. For the \textsf{PHOENIX} model we find $K_p = {163^{+49}_{-30}}$ \kms\ and $V_{\rm sys} = {-17^{+3}_{-2}}$ \kms. For Fe the values are ${K_p = {157^{+68}_{-44}}}$ \kms\ and ${V_{\rm sys} = {-16^{+2}_{-4}}}$ \kms, and for Fe$^+$ they are ${K_p = {178^{+41}_{-62}}}$ \kms\ and ${V_{\rm sys} = {-18^{+3}_{-3}}}$ \kms. Uncertainties correspond to the range of $K_p$ and $V_{\rm sys}$ within a $1\sigma$ contour around the peak. Because we only sample a small portion of the planet's orbit, only loose constraints on the semiamplitude $K_p$ are possible. The $V_{\rm sys}$ found here is offset by about $3$ \kms\ at the $\sim 1-2\sigma$ level. This blueshift may indicate winds in the upper atmosphere of the planet \citep{Kempton2012, CasasayasBarris2019}. Using values in Table~\ref{tab:par}, we predict a planetary RV semiamplitude of $K_p = 2\pi a \sin i_p/P = 217.4 \pm 6.2$ \kms. {This value is higher than the $K_p$ measured from cross-correlation, but still consistent to within the $1\sigma$ uncertainties.}

Equation~\ref{eqn:cc} involves a normalization term in the denominator that allows the CCF to return a physically meaningful quantity \citep{Hoeijmakers2019}. The CCF peak is a weighted average of the depths of individual lines in the transmission spectrum of the planet. In practice, the average depth depends on the weighting used for low S/N pixels ($w(i)$) and the wavelength range of the cross-correlation; it also does not correspond to the depth of any particular line. However, it provides an order-of-magnitude estimate of typical absorption depths, and hence the altitude of the species in the exoplanet's atmosphere. We refer to the average absorption depth as $\langle\delta\rangle$, which is equal to the peak value of the stacked CCF over all $K_p$ and $V_{\rm sys}$ combinations. As shown by \citet{Hoeijmakers2019}, Fe lines probe much deeper in the atmosphere than Fe$^+$ lines under chemical equilibrium. While Fe$^+$ lines are stronger in the optical, they are fewer in number; Fe$^+$ absorption is generally much stronger in the near ultraviolet \citep[e.g.][]{Sing2019}. We find average absorption depths of ${{(3.6\pm0.8)\times10^{-4}}}$ and ${{(1.5\pm0.4)\times10^{-3}}}$ for Fe and Fe$^+$ respectively (note, the significance of the Fe$^+$ signal only indicates evidence of the species, but we can still proceed with using the signal to learn about the planet). 

Per Equation~\ref{eqn:cc}, the average absorption depth depends on the absolute depths of lines in the data, as well as the relative (but not absolute) depths of lines in the model. The results above are of the same order of magnitude as those for KELT-9b \citep{Hoeijmakers2019}. The height of the atmosphere ($H$) extends 5--10 scale heights ($H_{\rm sc}$, of length hundreds of kilometers for hot Jupiters) \citep{Madhusudhan2014rev}. The excess absorption beyond the transmission spectrum continuum $(R_p/R_\star)^2$ is approximately $\delta \approx 2R_pH/R_\star^2$; in other words, $H \approx \delta R_p / 2(R_p/R_\star)^2$. For order of magnitude estimates, we use values in Table~\ref{tab:par} and assume the base of the atmosphere has a pressure of {0.01 bar \citep{Kitzmann2018}, which is typical for the H$^{-}$ continuum of an UHJ}. We also take $H_{\rm sc} \sim 880$ km, estimated from the measured $T_{\rm eq}$ and $\log g$, as well as taking the mean molecular weight as $\mu = 2.3$ for an H$_2$-dominated atmosphere; however $\mu$ may be affected by H$_2$ dissociation on the planet's dayside. While the mass is highly uncertain, we take the posterior median value of $1.4$ \mjup\ in order to estimate $\log g$. The resultant pressures corresponding to the absorption are ${P \sim {6\times10^{-4}}}$ bar for Fe and ${P \sim {2\times10^{-7}}}$ bar for Fe$^+$. Interestingly, the blueshift is similar between both Fe and Fe$^+$ signals, suggesting that high-velocity winds might be fairly consistent across various depths in the atmosphere. 

The 4000 K Fe model returns the highest-significance detection. The Fe detection significances are 4.2$\sigma$, 4.7$\sigma$, and {5.2}$\sigma$ for temperatures of 2000, 3000, and 4000 K, respectively. The cross-correlation signal also decreases significantly during the second half of transit. The Fe detection significance is 4.6$\sigma$ when using exposures from only the first half of the transit. It drops to 1--2$\sigma$ if only exposures from the second half are used. This variability could trace differential chemistry between the morning and evening terminators. For example, \citet{Ehrenreich2020} infer a lack of neutral Fe vapor on the dayside terminator of WASP-76b based on the changing Doppler shift of the cross-correlation peak in each of their exposures. \citet{Hoeijmakers2020} observe slightly stronger Fe absorption in the second half of a transit of MASCARA-2b, which they suggest could be due to different temperatures or chemistry between terminators. In the case of TOI-1518b, additional transits would help improve our confidence that the observed variability is indeed of physical origin.

\subsection{Temperature and Circulation}

From the stellar radius, we can use the values of $R_{p}/R_\star$ and $a/R_\star$ from our {photometric} analysis to straightforwardly compute the planet's radius and orbital semimajor axis: {$R_{p}=1.875\pm0.053$~\rjup\ and $a=0.0389\pm0.0011$~au}. We also utilize the stellar parameters from the SED fit to further characterize the planet's atmosphere. The relative flux of the planet $D$ in the TESS bandpass, assuming no reflected starlight (i.e., zero geometric albedo), is related to the hemisphere-averaged brightness temperature $T_{p}$ via the following relation \citep[e.g.,][]{shporer2017}:
\begin{equation}
\label{depth} D  = \left(\frac{R_{p}}{R_\star}\right)^{2}\frac{\int F_{\lambda}(T_{p})\tau(\lambda)\lambda d\lambda}{\int F_{\lambda}(T_{\mathrm{eff}})\tau(\lambda)\lambda d\lambda}.
\end{equation}
Here, the stellar and planetary flux spectra are given by $F_{\lambda}(T_{\mathrm{eff}})$ and $F_{\lambda}(T_{p})$, respectively, and $\tau(\lambda)$ is the transmission function of the TESS bandpass. For simplicity, we assume that the planet's emission spectrum is well-modeled by a blackbody function. 

For the stellar spectrum, following the technique described in \citet{wong2020kelt9}, we use PHOENIX stellar models \citep{Husser2013} and calculate the integrated stellar flux in the denominator of Equation~\eqref{depth} for a grid of stellar parameters in the vicinity of the values derived from the SED fit. We then construct an empirical polynomial function in $\left\lbrace T_{\mathrm{eff}},\mathrm{[Fe/H]},\log g\right\rbrace$ that smoothly interpolates these values. The planet's brightness temperature can then be fit for using an MCMC routine, with Gaussian priors for $T_{\mathrm{eff}}$, $\mathrm{[Fe/H]}$, $\log g$, and $R_{p}/R_\star$ derived from the SED and TESS light-curve fits.

We use the secondary eclipse depth and nightside flux (Table~\ref{tab:par}) to calculate the corresponding dayside and nightside brightness temperatures of TOI-1518b: {$T_d = 3237\pm59$~K and $T_n=1700^{+700}_{-1200}$~K.} The extremely high dayside temperature makes TOI-1518b among the hottest exoplanets hitherto discovered, comparable to other UHJs such as WASP-18b ($3100\pm49$~K; \citealt{wong2020year1}) and WASP-33b ($3105\pm95$~K; \citealt{vonessen2020}). 

We note that any reflected light off the dayside atmosphere (i.e., nonzero geometric albedo) would decrease the contribution of the planet's thermal emission to the measured secondary eclipse, resulting in a lower inferred dayside brightness temperature. However, at these high temperatures, all known condensate species are expected to be in the vapor phase across the dayside hemisphere, making reflective clouds unlikely \citep[e.g.,][]{helling2019}. This is supported by emission spectrum modeling of other UHJs spanning optical and thermal infrared wavelengths, which break the degeneracy between short-wavelength reflectivity and planetary thermal emission and indicate geometric albedos consistent with zero \citep[e.g.,][]{shporer2019,wong2020year1,wong2021year2}.

{In the broader context of atmospheric circulation, the measured dayside and nightside brightness temperatures reflect the amount of absorbed insolation and the efficiency of day--night heat transport. We can use the simple thermal balance model outlined in \citet{cowanagol2011} to simultaneously constrain the Bond albedo $A_{\mathrm{B}}$ and the recirculation efficiency $\epsilon$. In this parametrization, $\epsilon$ ranges from 0 (no recirculation) to 1 (uniform global temperature). To properly propagate the uncertainties on the stellar and orbital parameters, we use the methodology described in \citet{wong2020kelt9}. Due to the highly-uncertain nightside brightness temperature, we retrieve very poor constraints: $A_{B}<0.2$ ($2\sigma$) and $\epsilon=0.5\pm0.3$. Higher signal-to-noise is required to construct a more precise picture of the atmospheric heat budget. This may be achieved either by including additional visible-wavelength photometry of the system from the TESS Extended Mission or by obtaining full-orbit phase-curve observations at infrared wavelengths, where the planet--star contrast ratio is significantly higher.}

\section{Discussion and Conclusions} \label{sec:discussion}

As there have been only a handful of previous detections of iron in UHJs, TOI-1518b adds an important additional data point in our efforts to understand the dynamics and thermal structure in highly irradiated atmospheres. We make a few concluding remarks about the planet below, and then compare it to other recently characterized UHJs.

\setlength{\tabcolsep}{9.0pt}
\begin{table*}
\centering
\begin{tabular}{l r r r c r}
\hline
Planet & $T_{\rm eq}$ (K) & $R_p$ (\rjup) & $\log g$ (cgs) & {Fe (Transmission/Emission)} & Reference\\
\hline 
  TOI-1518b    & $2492\pm38$  & $1.875 \pm 0.053$         & $<3.229$                  & Y/- & this study \\
  KELT-9b      & $4050\pm180$ & $1.783 \pm 0.009$         & $3.30^{+0.11}_{-0.15}$    & Y/Y & H18, H19, {P20} \\
  MASCARA-2b   & $2260\pm50$  & $1.83\pm0.07$             & $<3.467$                  & Y/- & CB19, S20, H20 \\
  WASP-121b    & $2358\pm52$  & $1.865\pm0.044$           & $2.973\pm0.017$           & Y/- & D16, C19 \\
  WASP-76b     & $2228\pm122$ & $1.854\pm0.077$           & $2.806\pm0.034$           & Y/- & E20 \\
  WASP-189b    & $2641\pm34$  & $1.619\pm0.021$           & $3.274^{+0.048}_{-0.042}$ & -/Y & A18, C20, L20, {Y20} \\
  WASP-33b     & $2710\pm50$  & $1.679^{+0.019}_{-0.030}$ & $3.297^{+0.043}_{-0.041}$ & -/Y & Y19, {N20} \\
  {WASP-19b}  & $2372\pm60$ & $1.392\pm0.040$ & $2.616^{+0.065}_{-0.070}$ & -/- & {W16, Se21} \\
  {TOI-1431b} & $2181\pm95$ & $1.546\pm0.063$ & $4.148^{+0.043}_{-0.041}$ & -/- & {S21, A21} \\
  \hline
\end{tabular}
  \caption{Summary of recent high-resolution spectroscopy iron detections, comparing TOI-1518b to known transiting ultra-hot Jupiters. Values and uncertainties for equilibrium temperature and planet radius are reported in the references. Surface gravity was calculated from available parameters, if not reported explicitly. References: H18 \citep{Hoeijmakers2018}, H19 \citep{Hoeijmakers2019}, {P20 \citep{Pino2020}} CB19 \citep{CasasayasBarris2019}, H20 \citep{Hoeijmakers2020}, S20 \citep{Stangret2020}, D16 \citep{Delrez2016}, C19 \citep{Cabot2019}, E20 \citep{Ehrenreich2020}, A18 \citep{Anderson2018}, C20 \citep{Cauley2020}, L20 \citep{Lendl2020}, {Y20 \citep{Yan2020c}}, Y19 \citep{Yan2019}, {N20 \citep{Nugroho2020b}}, {W16 \citep{Wong2016}}, {Se21 \citep{Sedaghati2021}}, {S21 \citep{Stangret2021}}, {A21 \citep{Addison2021}}.}
  \label{tab:fepl}
\end{table*}

\subsection{TOI-1518b In the Context of Other Iron Detections} \label{subsec:otherFe}

Alkali metals (Na and K) have been detected in transmission for numerous hot Jupiters \citep[e.g.][]{Sing2016}. Over the past two years, Fe has also become an increasingly common detected species, albeit mostly in UHJs with $T_{\rm eq} \gtrsim 2000$ K \citep{Parmentier2018}. Fe traces winds in the upper atmosphere through the systemic velocity offset of the cross-correlation peak and is also a potential non-oxide contributor to thermal inversions \citep{Lothringer2018}. In the literature, Fe has been detected in transmission in the following exoplanets: KELT-9b \citep{Hoeijmakers2018, Hoeijmakers2019}, WASP-121b \citep{Cabot2019}, MASCARA-2b \citep{Stangret2020, Hoeijmakers2020}, WASP-76b \citep{Ehrenreich2020}, and TOI-1518b (this study). {Fe has been detected in emission in KELT-9b \citep{Pino2020}, WASP-189b \citep{Lendl2020}, and WASP-33b \citep{Yan2020c}}. These targets are listed in Table~\ref{tab:fepl}.

Interestingly, \citet{Cauley2020} do not detect Fe in transmission in WASP-189b, despite it being one of the brightest and hottest systems and the fact that Fe is detected in emission {(however, the observations were made under poor weather conditions)}. We note that, {although} Ca$^+$ {was found} in transmission in WASP-33b \citep{Yan2019}, {and} Fe in emission \citep{Yan2020c}, there has been no claim of Fe in transmission. {Fe may be especially difficult to detect in WASP-33b due to stellar pulsations}. We acknowledge {a few additional recent studies, including the non-detection Fe in WASP-19b \citep{Sedaghati2021} which is listed in Table~\ref{tab:fepl} (however this target is considerably fainter than the others, at $V=12.3$), a recent transmission spectroscopy study of HD149026b \citep{Ishizuka2021} (however the Fe signal was only at $2.8 \sigma$), and a non-detection in TOI-1431b (which orbits a relatively bright $V=8.0$ star; this target is listed in Table~\ref{tab:fepl}).}

While the statistical sample is small, Fe detections seem to favor particularly inflated UHJs, potentially with a cutoff around $1.7-1.8$ \rjup. One explanation is that Fe detections require particularly large atmospheric scale heights in order for the atoms to imprint sufficiently deep absorption lines on top of the continuum of the transmission spectrum. However, the surface gravity, which is inversely proportional to scale height, does not show a discernible relationship to Fe detections. For example, Fe was detected in transmission in KELT-9b, whose large mass yields a similar $\log g$ as WASP-189b. The $\log g$ of TOI-1518b is less than $3.229$ at $95\%$ confidence. {There are a few bright targets with $R_p < 1.7$ \rjup\ that are without detailed, cross-correlation atmospheric analyses, and do not have reported detections of Fe in transmission:} MASCARA-1b \citep{Talens2017}, KELT-7b \citep{Bieryla2015}, and KELT-17b \citep{Zhou2016}. As more gas giants are detected and characterized, it will be interesting to see if such a trend between Fe detection and planetary radius continues to hold.

\subsection{Photometric Mass Measurement and Caveats}\label{subsec:photmass}
In our analysis of the TESS photometry, we obtain a strong detection of the ellipsoidal distortion component of the phase-curve variability. This signal is driven by the tidal response of the stellar surface to the mutual star--planet gravitational interaction, which in turn depends on the mass ratio between the two components. It follows that the measured amplitude of the ellipsoidal distortion signal can be used to obtain an independent estimate of the planet's mass.

The ellipsoidal distortion of the star is formally modeled as a series of cosine terms, with the semiamplitude of the leading term (at the first harmonic of the orbital phase) related to fundamental parameters of the system via the following expression \citep[e.g.,][]{morris1985,shporer2017}:
\begin{equation}\label{ellip}
A_{\mathrm{ellip}}= \alpha_{\mathrm{ellip}}\frac{M_{p}}{M_\star}\left(\frac{R_\star}{a}\right)^{3}\sin^2 i_p.
\end{equation}
Here, the pre-factor $\alpha_{\mathrm{ellip}}$ is a function of the linear limb-darkening and gravity-darkening coefficients $u$ and $g$ for the host star:
\begin{equation}\label{alphaellip}
\alpha_{\mathrm{ellip}}=\frac{3}{20}\frac{(u+15)(g+1)}{3-u}.
\end{equation}

Similar to our treatment of the quadratic limb-darkening coefficients in the TESS phase-curve analysis (Section~\ref{subsec:phasecurve}), we construct Gaussian priors for $u$ and $g$ using values interpolated from the coefficients listed in \citet{claret2017}: $u=0.41\pm0.05$ and $g=0.12\pm0.05$. We then use Equations~\eqref{ellip} and \eqref{alphaellip} to construct the posterior for $M_{p}$ through Monte Carlo sampling of the distribution of values for $A_{\mathrm{ellip}}$, $a/R_\star$, $i_{p}$, $M_\star$, $u$, and $g$. {We obtain a photometric mass estimate of $M_{p}=4.8^{+1.3}_{-1.1}$~\mjup. This value is significantly ($2.3\sigma$)} larger than the RV-derived mass upper limit of 2.3 \mjup.

This discrepancy between the phase-curve-derived and RV-derived masses may be attributable to oversimplifications in the stellar tidal response formalism. \cite{Gomel2021} found a discrepancy of up to 30\% between the amplitudes of the ellipsoidal distortion derived from the analytic expressions of \cite{morris1985} and those derived numerically. More fundamentally, the classical theory of stellar ellipsoidal distortion from which Equations~\eqref{ellip} and \eqref{alphaellip} are derived makes several key assumptions: (1) steady-state approximation, which assumes that the star is in hydrostatic balance and ignores fluid inertia and the possibility of dynamical tides, (2) equatorial orbit of the companion, and (3) no effects from stellar rotation. The last two assumptions in particular are ostensibly invalid in the case of the TOI-1518 system, which contains a hot Jupiter on a misaligned orbit around a rapidly-rotating star (see Section~\ref{subsec:rm}). The fast rotation of the star and the resulting rotational bulge, combined with the spin-orbit misalignment, mean that the tidal bulge raised by the planet traverses regions of the stellar surface with significantly different surface gravities. This is expected to directly affect the tidal response of the star and the corresponding amplitude of the ellipsoidal distortion signal.

Another possible contributor to an unexpected first harmonic phase-curve modulation is the variable stellar irradiation experienced by the planet. This scenario was explored in detail for the case of KELT-9 --- a similarly misaligned system with an ultra-hot Jupiter around a rapidly-rotating star --- where it was found to be the primary source of the unusual phase alignment of the measured first harmonic photometric modulation \citep{wong2020kelt9}. In short, the rapid stellar rotation induces variations in the effective temperature of the planet-facing hemisphere, which cause the planetary thermal emission to change in response to the time-varying insolation. The three-dimensional orientation of TOI-1518's rotation axis is not known from the available data, preventing us from being able to directly model the relative phasing of this additional irradiation signal (as was done for the KELT-9 system). Nevertheless, we do expect some level of photometric variability at the first harmonic that is due to the planet's variable dayside temperature, which may bias the photometric mass estimate.

The previous discussion serves as a cautionary tale about the reliability of photometric mass measurements derived from the ellipsoidal distortion signal. The complexities of the stellar tidal response and the possibility of additional contributions from the planet's thermal emission mean that many systems are susceptible to significant discrepancies between the measured and expected first harmonic amplitudes. Future RV monitoring of this system will improve the precision of the planet's mass.

\subsection{Conclusion} \label{subsec:con}

TESS continues to find numerous transiting exoplanet candidates. As these planets are confirmed, some are bound to become interesting case studies for atmospheric characterization. In this paper, we reported the confirmation of an ultra-hot Jupiter on a close-in, highly misaligned orbit around TOI-1518. The stellar, planetary, and orbital parameters derived from fitting the TESS light curve{, ground-based transit photometry,} and spectral energy distribution are listed in Table~\ref{tab:par}. The photometry displays a clear secondary eclipse signal, as well as phase-synchronized modulations in flux attributed to the day--night brightness contrast of the planet and the tidal distortion of the host star. In addition, we searched for neutral and ionized Fe in the companion's atmosphere through high-resolution transmission spectroscopy. We detected Fe at high confidence, and also found evidence for Fe$^+$. TOI-1518b is highly inflated, which makes it amenable to intensive atmospheric characterization. The equilibrium temperature of TOI-1518b is in the regime where the planet might exhibit a thermal inversion \citep{Fortney2008, Lothringer2018, 2019Malik, Gandhi2019}. This, combined with the brightness of the host star, makes TOI-1518b an attractive target for follow-up emission spectroscopy \citep{Pino2020, Nugroho2020b, Yan2020c}. 

\acknowledgements

This paper includes data collected by the TESS mission. Funding for the TESS mission is provided by the NASA Explorer Program. Resources supporting this work were provided by the NASA High-End Computing (HEC) Program through the NASA Advanced Supercomputing (NAS) Division at Ames Research Center for the production of the SPOC data products. We acknowledge the use of TESS High Level Science Products (HLSP) produced by the Quick-Look Pipeline (QLP) at the TESS Science Office at MIT, which are publicly available from the Mikulski Archive for Space Telescopes (MAST). 
This work used data from the EXtreme PREcision Spectrograph (EXPRES) that was designed and commissioned at Yale with financial support by the U.S. National Science Foundation under MRI-1429365 and ATI1509436 (PI D. Fischer). We gratefully acknowledge support for telescope time using EXPRES at the LDT from the Heising-Simons Foundation and an anonymous
Yale donor. We acknowledge support from U.S. National Science Foundation grant 2009528.
This research made use of Lightkurve, a Python package for Kepler and TESS data analysis (Lightkurve Collaboration, 2018). This paper is partially based on observations made with the Nordic Optical Telescope, operated by the Nordic Optical Telescope Scientific Association at the Observatorio del Roque de los Muchachos, La Palma, Spain, of the Instituto de Astrofisica de Canarias.
K.K.M. gratefully acknowledges support from the New York Community Trust's Fund for Astrophysical Research.
I.W. is supported by a Heising-Simons \textit{51 Pegasi b} postdoctoral fellowship. 
A.A.B., B.S.S. and I.A.S. acknowledge the support of Ministry of Science and Higher Education of the Russian Federation under the grant 075-15-2020-780 (N13.1902.21.0039). This paper is partially based on observations made at the CMO SAI MSU with the support by M.V. Lomonosov Moscow State University Program of Development. VA was supported by a research grant (00028173) from VILLUM FONDEN. Funding for the Stellar Astrophysics Centre is provided by The Danish National Research Foundation (Grant agreement no.: DNRF106).
This research made use of \textsf{exoplanet} \citep{exoplanet:exoplanet} and its
dependencies \citep{exoplanet:agol20, exoplanet:arviz, exoplanet:astropy13,
exoplanet:astropy18,  exoplanet:luger18, exoplanet:pymc3,
exoplanet:theano}. {We also acknowledge very useful input from an anonymous referee, which improved the clarity and structure of the manuscript}.

\software{Tapir \citep{Jensen:2013}, AstroImageJ \citep{Collins:2017}, molecfit \citep{smette2015}, radvel \citep{Fulton2018}, Lightkurve \citep{Lightkurve2018}, Helios-K \citep{2021Grimm}, FastChem \citep{2018Stock}, batman \citep{batman}, emcee \citep{Foreman-Mackey2013}, SME \citep{Valenti1996}, exoplanet \citep{exoplanet:exoplanet}, astropy \citep{exoplanet:astropy18}}

\bibliography{refs}{}
\bibliographystyle{aasjournal}

\begin{figure}[b!]
    \centering
    \includegraphics[width=0.9\linewidth]{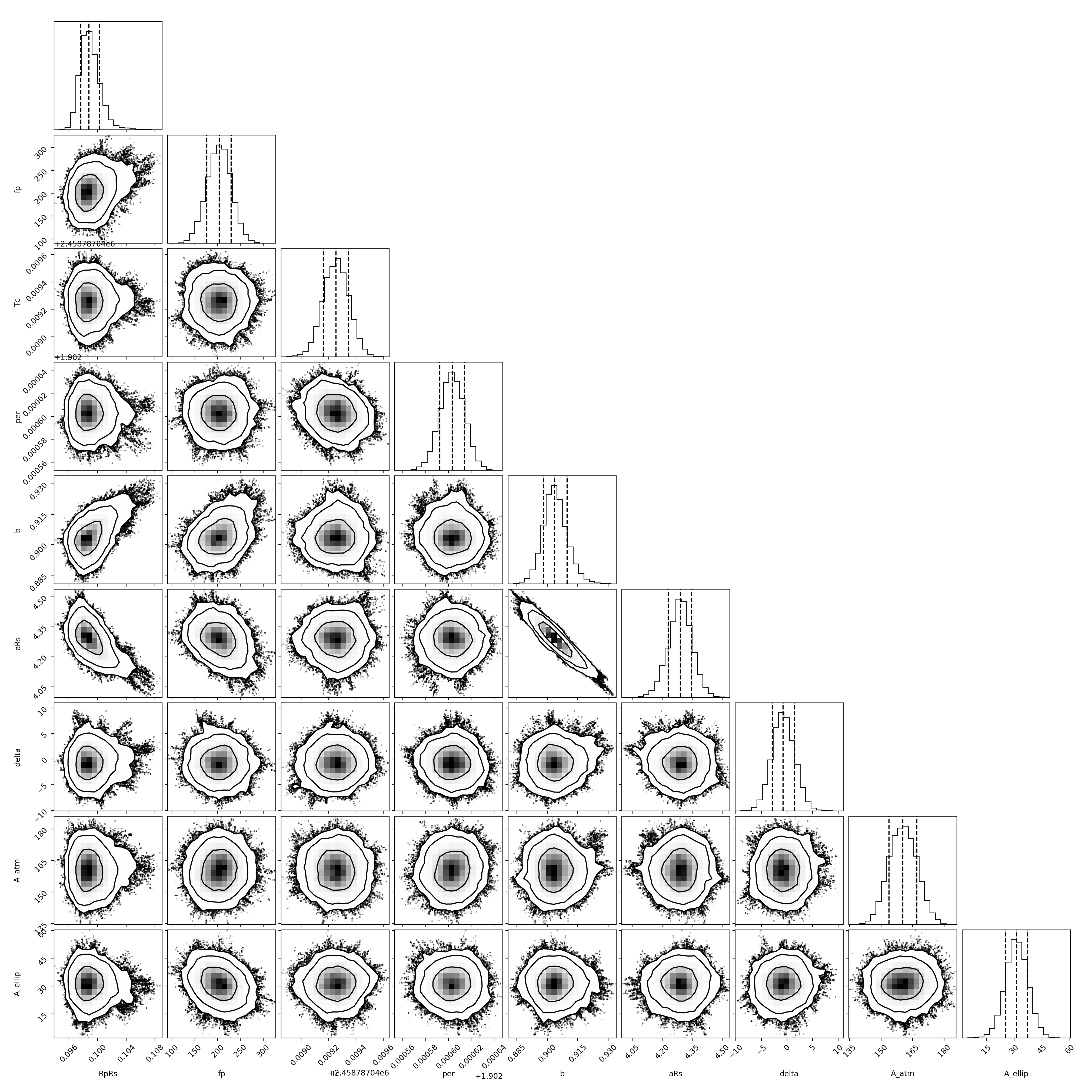}
    \caption{{Corner plot of parameters involved in the {joint TESS and ground-based light-curve fit}.}}
    \label{fig:corner}
\end{figure}

\appendix
{Figure~\ref{fig:corner}} depicts a corner plot containing all the astrophysical parameters fitted {for in our joint analysis of the TESS light curve and ground-based full-transit photometry (Section~\ref{subsec:joint}); for clarity, the limb-darkening coefficients for each dataset are not shown.} The values of the average relative planetary flux $f_p$, planetary atmospheric brightness modulation amplitude $A_{\mathrm{atm}}$, and stellar ellipsoidal distortion amplitude $A_{\mathrm{ellip}}$ are given in parts-per-million. The phase offset in the planetary phase curve $\delta$ is provided in degrees. Note the significant correlations between the impact parameter $b$, scaled semimajor axis $a/R_\star$, radius ratio $R_p/R_\star$, and $f_p$ --- a consequence of the grazing nature of the planetary transit.

\begin{figure}
    \centering
    \includegraphics[width=\linewidth]{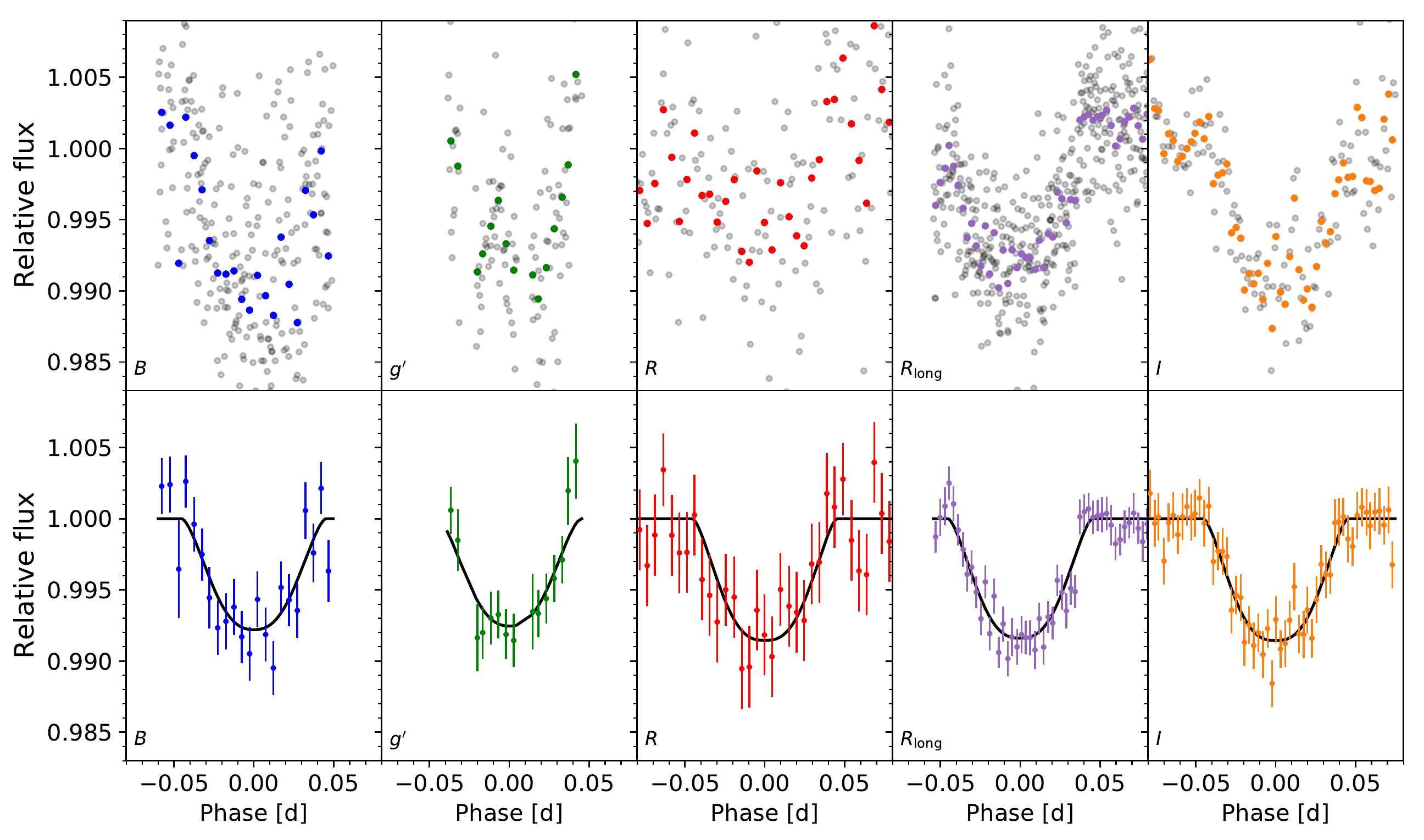}
    \caption{{Ground-based light curves of TOI-1518 with full coverage of the primary transits, collected as part of the TESS Follow-up Program. Top: the photometry at the native time resolution (gray points) and binned (colored points). Each panel is labeled by the respective bandpass. The binning interval for the $B$-, $g'$-, and $R$-band observations is 7 minutes; a shorter 4-minute bin size is used for the higher-precision $R_{\rm long}$- and $I$-band transits. Bottom: binned, systematics-corrected light curves, with the best-fit transit model from the joint TESS and ground-based photometric fit (Table~\ref{tab:par}) plotted in black.}}
    \label{fig:groundphot}
\end{figure}

{Figure~\ref{fig:groundphot} shows the full-transit light curves collected as part of ground-based followup observations, as described in Section \ref{subsec:groundlc}. Each light curve is labeled with the filter used. The best-fit transit model from the joint TESS and ground-based photometric fit is shown in the bottom panels.} 

\end{document}